\documentclass[aps,prd,reprint,superscriptaddress,longbibliography]{revtex4-2}
\usepackage{amsmath,amssymb,amsthm,mathtools,bm,mathrsfs}
\usepackage{graphicx,booktabs,microtype}
\usepackage[hidelinks]{hyperref}
\graphicspath{{figures/}{./}}
\newcommand{\dd}{\mathrm d}
\newcommand{\bk}{\boldsymbol k}
\newcommand{\bp}{\boldsymbol p}
\newcommand{\bq}{\boldsymbol q}
\newcommand{\bb}{\boldsymbol b}
\newcommand{\kt}{k_T}

\newcommand{\KCSS}{\mathcal K_{\mathrm{CSS2}}}

\newcommand{\NP}{\mathrm{NP}}
\newcommand{\pert}{\mathrm{pert}}
\newcommand{\LP}{\mathrm{LP}}
\newcommand{\FO}{\mathrm{FO}}
\newcommand{\Nthree}{N^3\mathrm{LL}'}
\newcommand{\ot}{\otimes_T}
\newcommand{\ox}{\otimes_x}
\newcommand{\deltaT}{\delta^{(2)}}
\newcommand{\Ln}{\mathcal L}
\newcommand{\Jdist}{\mathfrak L}
\newcommand{\Ggen}{\mathcal G}
\newcommand{\Ccal}{\mathcal C}
\newcommand{\Dcal}{\mathcal D}
\newcommand{\Uop}{\mathbb U}
\newcommand{\Dop}{\mathbb D}
\newcommand{\FT}{\mathcal F_{k\to b}}
\newcommand{\IFT}{\mathcal F^{-1}_{b\to k}}
\newcommand{\Sch}{\mathcal S(\mathbb R^2)}
\newcommand{\Schp}{\mathcal S'(\mathbb R^2)}
\newcommand{\Ltwo}{L^2(\mathbb R^2)}
\newcommand{\Dlog}{\mathscr D_{\log}}
\newcommand{\Order}[1]{\mathcal O\!\left(#1\right)}
\newcommand{\pair}[2]{\langle #1,#2\rangle}
\newcommand{\norm}[1]{\lVert #1\rVert}
\newcommand{\ExpT}{\operatorname{Exp}_{\otimes_T}}
\DeclareMathOperator{\Dom}{Dom}

\theoremstyle{plain}
\newtheorem{theorem}{Theorem}
\newtheorem{proposition}[theorem]{Proposition}

\newtheorem{corollary}[theorem]{Corollary}
\theoremstyle{definition}

\theoremstyle{remark}

\begin{document}
\title{Native transverse-momentum evolution and matching for TMD factorization and extraction}

\author{D.\ Keller}
\email{dustin@virginia.edu}
\affiliation{Department of Physics, University of Virginia, Charlottesville, Virginia 22904, USA}

%\date{Review draft, 1 October 2026}
\begin{abstract}
We present a native transverse-momentum realization of the Collins/CSS2
leading-power transverse-momentum-dependent (TMD) factorization framework.
The construction connects fixed-order distributional matching, origin contacts,
unexpanded two-scale evolution, and a differentiable interface for hadronic
structure. An endpoint-independent canonical scale with a specified infrared
completion generates a Collins--Soper kernel on a common invariant domain.
Complete canonical legs are bounded and admit constructive momentum-space
evaluation through resolvent convolutions or elliptic solves. We give contact
and finite-bin error bounds, parent-to-momentum-boundary translations, ordered
flavor recurrences, and an all-order argument for the four exponent towers of
$N^3\mathrm{LL}$ evolution. Localized joint-$(x,k_T)$ inputs, nonperturbative
rapidity broadening, regulated moments and the full-spectrum subtraction are
formulated as distinct parts of the prescription. The physical realization uses
sourced four-loop beta/cusp and three-loop noncusp/Collins--Soper ingredients
with strict-NLO hard/OPE boundaries. Non-Gaussian continuum tests and three-scale
physical-PDF comparisons validate ordinary legs, contacts, individual insertions
and analytic input derivatives. Including a declared bound on the continuation
above a finite PDF table, physical bin discrepancies remain below
$4.27\times10^{-5}$ of the peak reference bin. A qualified proton--proton
Drell--Yan and neural-closure demonstration connects the method to inverse use.
Higher-order boundary assembly and matched observables are specified without
claiming an implemented $N^3\mathrm{LL}'$ spectrum or a completed global extraction.
\end{abstract}
\maketitle
\raggedbottom
\section{Introduction}
\label{sec:introduction}

The transverse motion of quarks and gluons carries information about hadron structure that is integrated out of collinear parton distributions. Measurements of Drell--Yan recoil, semi-inclusive deep-inelastic scattering, and nearly back-to-back hadrons resolve this structure through convolutions of transverse-momentum-dependent distributions (TMDs). In the region of measured transverse momentum $q_T\ll Q$, the factorization framework separates the hard reaction, short-distance matching, rapidity evolution, and long-distance transverse dynamics~\cite{CollinsSoper1981,CSS1985,Collins2011,TMDHandbook2023}. Precision extractions must describe both the perturbative radiation that broadens the spectrum and the hadronic information that remains after that radiation has been accounted for.

This problem is particularly attractive for differentiable inverse methods. A neural network can represent correlated longitudinal and transverse dependence, while an integral forward map connects its output to cross sections. Fernando and Keller demonstrated this strategy with E288 and E605 fixed-target Drell--Yan data~\cite{FernandoKellerMomentum2026}. Their two-stage procedure learns a cross-section-level transverse kernel and reconstructs a normalized momentum-space profile through differentiable quadrature. The construction keeps the fit in transverse momentum and exposes correlated dependence on the longitudinal fraction and hard scale. It provides a concrete motivation for embedding a high-accuracy perturbative evolution operator in the same computational representation.

The proposed advantage is one of organization of the inverse problem. A momentum-space parameterization permits regularity, localization, and input normalization to be imposed directly on the function being inferred. The observable map retains the measured recoil variable, its binning, and the transverse convolution relating it to partonic momenta. A network can then express correlations supported by the data without imposing a separable longitudinal--transverse ansatz. The measured $q_T$ nevertheless constrains convolutions of the partonic $k_T$ distributions: working in momentum space preserves this physical relation rather than identifying these two variables. The amount of recoverable information is determined by the experimental coverage and the sensitivity of the forward operator.

Perturbative control is central to this program. In a precision extraction, the scale dependence fixed by QCD should enter through the evolution equations and matching coefficients, while the fitted flexibility describes the admitted long-distance input. Modern impact-parameter-space analyses provide demanding benchmarks, including high-logarithmic-accuracy global TMD fits and precision Drell--Yan spectra~\cite{MoosScimemiVladimirovZurita2024,MoosScimemiVladimirovZurita2025,BillisMichelTackmann2025}. Neural parameterizations have also been developed in that setting~\cite{BacchettaEtAl2025NN}. The practical question addressed here is how to retain this perturbative information while acting directly on a momentum-space input throughout the extraction.

Transverse-momentum resummation has a substantial direct-space literature. Early approaches to electroweak-boson spectra~\cite{EllisVeseli1998,KuleszaStirling1999}, modern high-order momentum-space resummation~\cite{MonniReTorrielli2016,BizonEtAl2018}, distribution-space renormalization-group evolution~\cite{EbertTackmann2017}, and hybrid semi-analytic methods~\cite{KangLeeVaidya2018} have established complementary ways to organize the recoil logarithms. Recent analytic momentum-space work~\cite{Simonelli2026} and studies of infrared-prescription dependence~\cite{CeruttiSimonelli2026} sharpen the interest in making the perturbative operator and its long-distance completion explicit. The present construction concerns the TMD-level operator used inside an extraction: its domain, its unexpanded two-scale evolution, its constructive momentum-space evaluation, and its logarithmic accuracy.

We take Collins/CSS2 as the parent operator convention~\cite{CollinsRogers2017}. Its renormalized TMD definitions, soft subtraction, Wilson-line process class, and leading-power factorization assumptions are retained. We denote their momentum-space representation by $\KCSS$. The central construction uses functions of the nonnegative self-adjoint operator $T=-\nabla_k^2$. The scalar canonical CSS2 solution becomes an operator acting on momentum-space functions; a sequence of resolvent convolutions evaluates that operator without numerical Fourier or Hankel transformations of the fitted input. Fourier duality supplies the proof of representation equivalence, while the numerical operations are carried out in momentum space.

Three analytic results make this an explicit resummed prescription. First, an infrared-completed canonical scale generates an unexpanded Collins--Soper kernel whose endpoint evolution exists on a common invariant domain. Second, the complete canonical leg is bounded and admits a convergent momentum-space evaluator that preserves its contact term. Third, the retained renormalization-group functions reproduce the four exponent towers required at $N^3\mathrm{LL}$. The physical realization tested here uses strict-NLO hard and OPE boundaries. We also give the order-by-order assembly and boundary translations needed for a future primed-order implementation; physical three-loop hard/matching ingestion is not a result of the present calculation. These results are established for a fixed-flavor massless system, a specified infrared completion, and longitudinal input functions obeying the growth assumptions given below.

For phenomenology, this produces the architecture
\begin{equation}
 S_{i/h,\theta}(x,\bk)
 \ \longmapsto\ F^c_{i/h}(x,\bk;\mu,\zeta)
 \ \longmapsto\ W^c(\bq)
 \ \longmapsto\ \sigma_{\mathcal B}(\theta),
 \label{eq:architecture}
\end{equation}
where $S_{i/h,\theta}$ is an admitted long-distance input and the intervening operators carry perturbative matching, evolution, and the physical measurement. A deep neural network (DNN) is one realization of $S_{i/h,\theta}$; analytic functions, splines, and mixtures use the same perturbative construction. Boundedness of the canonical operator also gives a controlled way to differentiate the forward map and its approximations.

The numerical results establish more than a distributional change of notation.
The same uncapped canonical prescription is evaluated with physical PDFs and
strict-NLO boundary components at three scales, including non-Gaussian two-leg
inputs, contacts and analytic derivatives. The maximum peak-normalized bin
error, including a stated ultraviolet PDF continuation bound, is below
$4.27\times10^{-5}$. A restricted neural closure supplies a separate inverse
demonstration. The general joint-$(x,k_T)$ fitting interface is part of the
method; an identical-data comparison with the later $b_T$-space extraction
\cite{FernandoKeller2026} is reserved for a subsequent phenomenological study.

The distinction from earlier direct-space work is therefore the complete
single-TMD operator construction on a stated domain, its exact contact
extension, and its action on an arbitrary admitted momentum-space input.
The parent factorization is imported, and no new factorization theorem or
priority claim for momentum-space resummation is made. The prescription,
representation and numerical evaluator are specified separately so that each
can be compared with an alternative without changing the other ingredients.

The paper is organized around this physics application. Section~\ref{sec:factorization} defines the factorization and scale conventions. Section~\ref{sec:fixedorder} develops the fixed-order distribution calculus and boundary matching. Sections~\ref{sec:unexpanded} and \ref{sec:evaluation} construct and evaluate the admitted unexpanded operator. Its logarithmic accuracy is proved in section~\ref{sec:accuracy}. Sections~\ref{sec:extraction} and \ref{sec:observables} give the long-distance and measurement interfaces, including differentiable fitting, regulated moments, and full-spectrum matching. Section~\ref{sec:verification} presents the source-qualified implementation and continuum validation, and Sec.~\ref{sec:demo} gives a bounded inverse demonstration. The appendices collect explicit perturbative kernels and proofs needed for the distributional interfaces. The companion scientific software and reproduction instructions are maintained in the public \texttt{k-space} repository~\cite{KSpaceSoftware}.

\section{Factorization and the momentum-space extraction operator}
\label{sec:factorization}

\subsection{Parent definitions and scale conventions}

Let $\bk$, $\bq$, and $\bb$ denote transverse vectors, with magnitudes $k_T$, $q_T$, and $b_T$. We use
\begin{align}
 \widetilde F(x,\bb;\mu,\zeta)
   &=\int\dd^2\bk\,e^{i\bb\cdot\bk}F(x,\bk;\mu,\zeta),\notag\\
 F(x,\bk;\mu,\zeta)
   &=\int\frac{\dd^2\bb}{(2\pi)^2}
              e^{-i\bb\cdot\bk}\widetilde F(x,\bb;\mu,\zeta).
 \label{eq:fourier}
\end{align}
The transverse and longitudinal convolutions are
\begin{align}
 (A\ot B)(\bq)&=\int\dd^2\bk\,A(\bk)B(\bq-\bk),\notag\\
 (C\ox f)(x)&=\int_x^1\frac{\dd z}{z}\,C(z)f(x/z).
 \label{eq:convolutions}
\end{align}
Singular transverse convolutions are defined on the domains specified in section~\ref{sec:fixedorder}. The coupling and logarithms are
\begin{equation}
\begin{aligned}
 a(\mu)&=\frac{\alpha_s(\mu)}{4\pi},&
 L_b&=\ln\frac{\mu^2b_T^2}{b_0^2},\\
 \ell_\zeta&=\ln\frac{\zeta}{\mu^2},&
 b_0&=2e^{-\gamma_E}.
\end{aligned}
 \label{eq:conventions}
\end{equation}
Our Collins--Soper convention is $D_i=-\widetilde K_i^{\rm Collins}$, so that
\begin{align}
 \partial_{\ln\sqrt\zeta}\widetilde F_i&=-D_i\widetilde F_i,\notag\\
 \partial_{\ln\mu}\widetilde F_i
  &=\gamma_F^i(a,\ell_\zeta)\widetilde F_i,
 &\gamma_F^i(a,\ell_\zeta)&=\gamma_i(a)-\Gamma_i(a)\ell_\zeta,\notag\\
 \frac{\dd D_i}{\dd\ln\mu}&=2\Gamma_i(a),
 &\frac{\dd a}{\dd\ln\mu}&=-2\sum_{r\ge0}\beta_r a^{r+2}.
 \label{eq:CSS-equations}
\end{align}
Here $\Gamma_i$ is the lightlike cusp anomalous dimension in the partonic representation $i$. These signs and factors follow the CSS2 convention in ref.~\cite{CollinsRogers2017}. In particular,
\begin{equation}
 \partial_{\ln\sqrt\zeta}\gamma_F^i=-2\Gamma_i,
 \label{eq:integrability}
\end{equation}
which fixes the flatness of the two-scale evolution connection.

The small-distance operator product expansion (OPE) is
\begin{equation}
\begin{aligned}
 \widetilde F_{i/h}(x,\bb;\mu,\zeta)
 &=\sum_j[\widetilde C_{i\leftarrow j}(\bb;\mu,\zeta)\ox f_{j/h}(\mu)](x)\\
 &\quad+\Order{(b_T\Lambda_{\rm QCD})^p}.
\end{aligned}
 \label{eq:ope}
\end{equation}
in its short-distance domain, with the omitted operator content fixing $p$~\cite{Collins2011,TMDHandbook2023}. A sufficient class for using this relation inside regulated moments is stated in appendix~\ref{app:inputs}. The high-order coefficient functions used by the construction are supplied in the same renormalization, rapidity, and flavor conventions.

\subsection{Drell--Yan and a generic measurement}

For a color-singlet Drell--Yan system of invariant mass $Q$, rapidity $y$, and hadronic center-of-mass energy $\sqrt{s_{\rm had}}$, the leading-power Born fractions are
\begin{equation}
 x_A=\frac{Q}{\sqrt{s_{\rm had}}}e^y,\qquad
 x_B=\frac{Q}{\sqrt{s_{\rm had}}}e^{-y}.
 \label{eq:bornfractions}
\end{equation}
The singular distribution has the form
\begin{equation}
\begin{aligned}
 W_{AB}(\bq;Q,y)
 &=\sum_q e_q^2\widehat H_{q\bar q}(Q,\mu)\\
 &\quad\times
 \bigl[F_{q/A}\ot F_{\bar q/B}
       +F_{\bar q/A}\ot F_{q/B}\bigr](\bq).
\end{aligned}
 \label{eq:dy}
\end{equation}
The suppressed arguments of the two legs are $(x_A;\mu,\zeta_A)$ and $(x_B;\mu,\zeta_B)$, respectively, with $\zeta_A\zeta_B=Q^4$. The reduced hard factor obeys $\widehat H^{(0)}=1$; the electromagnetic charges are displayed once, outside it. Overall Born normalization is denoted by $\sigma_0$ when forming the physical cross section. This convention cleanly separates the universal TMD legs from the process hard function.

A measured bin is a linear functional of the fully assembled cross section,
\begin{equation}
\begin{aligned}
 \sigma_{\mathcal B}
 &=\mathcal B[\dd\sigma_W+\dd\sigma_Y+\dd\sigma_{\rm pow}],\\
 \frac{\dd\sigma_W}{\dd\Phi\,\dd^2\bq}
 &=\sigma_0(\Phi)W(\Phi,\bq).
\end{aligned}
 \label{eq:measurement}
\end{equation}
The symbol $\mathcal B$ includes the stated phase-space integrals, acceptance, and recoil definition. A process whose TMD factorization contains two transverse legs can be written more generally as
\begin{equation}
 W_F=\sum_{ij}H_{ij\to F}\,
                 \mathcal M_F[F_{i/A},F_{j/B}],
 \label{eq:generic-process}
\end{equation}
where $\mathcal M_F$ includes the relevant bilinear transverse measurement. For SIDIS and back-to-back hadrons it also includes fragmentation momentum rescalings and Jacobians. Gluon channels may require tensor TMDs and angular projectors. The explicit proofs below address the radial scalar channel; tensor and fragmentation applications are specified by their own parent factorization, matching, and measurement data~\cite{EbertMistlbergerVita2021FF,LuoYangZhuZhu2021,ZhuLinearlyPolarized2026}.

\subsection{What is fitted and what is evolved}

The extraction interface used in this paper separates a hadron-dependent localized input from its short-distance operator:
\begin{equation}
 F_{i/h}^c(x;\mu,\zeta)
    =M_{i/h}(x,T;\mu,\zeta)S_{i/h,\theta}(x),
       \qquad T=-\nabla_k^2.
 \label{eq:central-leg}
\end{equation}
The superscript $c$ denotes the specified canonical completion. Section~\ref{sec:unexpanded} defines $M$ from canonical matching, collinear evolution, and the unexpanded CSS2 solution. The same $S_{i/h,\theta}$ is used at every final scale. This transfers the perturbatively predicted scale dependence from an unconstrained learned profile to the QCD forward operator. Hadron and flavor dependence remain available in the input; the CS kernel is common to its partonic universality class.

Equation~\eqref{eq:central-leg} also explains the main numerical distinction. A function of $T$ is nonlocal in $\bk$: the canonical scale belongs to the spectrum of a differential operator rather than to a pointwise replacement $\mu=k_T$. Its spectral variable has dimensions of inverse momentum squared. The scalar spectral nodes retain the resolution information associated with the parent $b_T^2$ variable, while the input and all operator applications remain represented in $k_T$. This distinction makes the transform-free evaluation precise.

\section{Fixed-order distribution calculus and matching}
\label{sec:fixedorder}

\subsection{An algebra with a specified domain}

The natural test space is $\Sch$, with tempered-distribution dual $\Schp$~\cite{Hormander1990}. At fixed perturbative order the transverse logarithms generate the closed algebra
\begin{equation}
 \mathscr A_\mu^{\rm pol}=\IFT\mathbb C[L_b],\qquad
 A\ot B:=\IFT[\widetilde A\widetilde B].
 \label{eq:polynomial-algebra}
\end{equation}
Finite polynomials in $L_b$ and their products are locally integrable at $b_T=0$ and have controlled growth at infinity. Equation~\eqref{eq:polynomial-algebra} therefore defines an associative, commutative product with unit $\deltaT$. It agrees with the usual transverse integral whenever that integral exists. Perturbative series are interpreted coefficient by coefficient in $\mathscr A_\mu^{\rm pol}[[a]]$.

For complete resummed multipliers we require each product used in a factorized expression to be admitted separately. This avoids extending the polynomial algebra beyond its domain. For instance, $e^{-3L_b/4}$ is locally integrable in two transverse dimensions, whereas its square requires an additional distributional extension at the origin. Section~\ref{sec:unexpanded} supplies a domain for the actual unexpanded evolution and section~\ref{sec:evaluation} covers its contact limits.

For $n\ge0$, the conventional radial plus distribution is fixed by
\begin{equation}
 \pair{\Ln_n(\mu)}{\varphi}
  =\int\frac{\dd^2\bk}{\pi k_T^2}
       \ln^n\!\frac{\mu^2}{k_T^2}
       \bigl[\varphi(\bk)-\varphi(\boldsymbol0)\theta(\mu-k_T)\bigr].
 \label{eq:plus-definition}
\end{equation}
Its canonical-disk integral is zero. With $L_r=\ln(\mu^2/q_r^2)$, its bin actions are
\begin{align}
 \mathcal B_n(q_a,q_b;\mu)
 &=\frac{L_a^{n+1}-L_b^{n+1}}{n+1},&&q_a>0,\notag\\
 \mathcal B_n(0,q_b;\mu)&=-\frac{L_b^{n+1}}{n+1}.
 \label{eq:plus-bins}
\end{align}
The first bin contains the support of $\deltaT$ by convention. These formulas are cumulative extensions to bin indicators; an arbitrary tempered distribution need not act on a sharp bin.

An analytic generator organizes the contact terms:
\begin{equation}
\begin{aligned}
 \FT[\Ggen_\eta]&=e^{-\eta L_b},\qquad 0<\operatorname{Re}\eta<1,\\
 \Ggen_\eta(\bk;\mu)
 &=\frac{e^{-2\gamma_E\eta}\Gamma(1-\eta)}{\pi\mu^2\Gamma(\eta)}
 \left(\frac{\mu^2}{k_T^2}\right)^{1-\eta}.
\end{aligned}
 \label{eq:generator}
\end{equation}
Its continuation near $\eta=0$ defines
\begin{equation}
\begin{aligned}
 \Jdist_n&=(-1)^{n+1}
 \left.\partial_\eta^{n+1}\Ggen_\eta\right|_{\eta=0},\\
 \FT[\Jdist_n]&=L_b^{n+1},\qquad \Jdist_{-1}=\deltaT .
\end{aligned}
 \label{eq:derivative-basis}
\end{equation}
The sign in the generator is tied to the convention $\ln(\mu^2/k_T^2)$ in eq.~\eqref{eq:plus-definition}. In particular,
\begin{equation}
 \Jdist_0=-\Ln_0,\qquad
 \Jdist_1=-2\Ln_1,\qquad
 \Jdist_2=-3\Ln_2-4\zeta_3\deltaT.
 \label{eq:first-dictionary}
\end{equation}
Here and below $\zeta_n$ denotes a Riemann-zeta value, distinct from the rapidity scale $\zeta$.

\begin{proposition}[Generator and polynomial composition]
\label{prop:generator}
For $\operatorname{Re}\eta>0$, $\operatorname{Re}\xi>0$, and $\operatorname{Re}(\eta+\xi)<1$, the ordinary transverse convolution gives
\begin{equation}
 \Ggen_\eta\ot\Ggen_\xi=\Ggen_{\eta+\xi}.
 \label{eq:generator-law}
\end{equation}
Coefficientwise continuation at the origin yields
\begin{equation}
 \Jdist_m\ot\Jdist_n=\Jdist_{m+n+1},\qquad
 \frac{\dd\Jdist_n}{\dd\ln\mu}=2(n+1)\Jdist_{n-1}.
 \label{eq:derivative-algebra}
\end{equation}
\end{proposition}
\begin{proof}
In the common strip, Feynman parametrization of the convergent massless integral gives
\begin{align}
 &\int\dd^2\bp\,(p_T^2)^{\eta-1}[(\bk-\bp)^2]^{\xi-1}\notag\\
 &\quad=\pi(k_T^2)^{\eta+\xi-1}
 \frac{\Gamma(\eta)\Gamma(\xi)\Gamma(1-\eta-\xi)}
      {\Gamma(1-\eta)\Gamma(1-\xi)\Gamma(\eta+\xi)}.
 \label{eq:riesz-integral}
\end{align}
The normalizations in eq.~\eqref{eq:generator} give eq.~\eqref{eq:generator-law}. Equivalently, the admitted Fourier multipliers multiply. Differentiating their holomorphic family near zero proves the polynomial convolution identity. Finally, differentiating $L_b^{n+1}$ with respect to $\ln\mu$ gives $2(n+1)L_b^n$; Fourier injectivity proves the scale derivative, including $n=0$.
\end{proof}

This is the distribution-space calculus underlying the fixed-order checks~\cite{EbertTackmann2017}. Its exponential form is
\begin{equation}
 \FT\!\left[\ExpT\!\left\{\sum_{n\ge0}r_n\Jdist_n\right\}\right]
       =\exp\!\left[\sum_{n\ge0}r_nL_b^{n+1}\right].
 \label{eq:formal-exponential}
\end{equation}
Equation~\eqref{eq:formal-exponential} is an identity of formal perturbative series; for an unexpanded expression its use additionally requires existence of the complete multiplier. We establish this for the RG-improved canonical construction in section~\ref{sec:unexpanded}.

\subsection{Coefficient import and the two canonical boundaries}

A parent matching coefficient at order $a^n$ has the form
\begin{equation}
 \widetilde C^{(n)}_{i\leftarrow j}(z,L_b,\ell_\zeta)
    =\sum_{m=0}^{2n}c^{(n,m)}_{ij}(z;\ell_\zeta)L_b^m.
 \label{eq:C-polynomial}
\end{equation}
Linearity and eq.~\eqref{eq:derivative-basis} give the exact fixed-order projection
\begin{equation}
\begin{aligned}
 \Ccal^{(n)}_{i\leftarrow j}(z,\bk;\mu,\zeta)
 &=c^{(n,0)}_{ij}(z;\ell_\zeta)\deltaT\\
 &\quad+\sum_{m=1}^{2n}c^{(n,m)}_{ij}(z;\ell_\zeta)
                     \Jdist_{m-1}(\bk;\mu).
\end{aligned}
 \label{eq:coefficient-map}
\end{equation}
Longitudinal coefficient distributions retain their $z$-space plus prescriptions. Known NNLO and N$^3$LO matching calculations supply the finite physical functions~\cite{EchevarriaScimemiVladimirov2016,EbertMistlbergerVita2020,LuoYangZhuZhu2021}. Their translation into the declared CSS2 convention precedes this transverse projection.

There are two useful notions of canonical boundary. The \emph{parent canonical boundary} is the coefficient at $L_b=\ell_\zeta=0$. The \emph{momentum-disk boundary} is the action on the disk $k_T\le\mu$. The finite translation between them is determined by
\begin{align}
 h(\eta)&=e^{-2\gamma_E\eta}\frac{\Gamma(1-\eta)}{\Gamma(1+\eta)},\notag\\
 \int_{k_T\le\mu}\dd^2\bk\,\Ggen_\eta&=h(\eta),\qquad
 J_m=(-1)^m h^{(m)}(0),\notag\\
 (J_0,J_1,\ldots,J_6)&=(1,0,0,-4\zeta_3,0,-48\zeta_5,160\zeta_3^2).
 \label{eq:boundary-moments}
\end{align}
Thus at zero rapidity logarithm,
\begin{align}
 c_{k}^{(2)}&=c^{(2,0)}-4\zeta_3c^{(2,3)},\notag\\
 c_{k}^{(3)}&=c^{(3,0)}-4\zeta_3c^{(3,3)}
                   -48\zeta_5c^{(3,5)}+160\zeta_3^2c^{(3,6)}.
 \label{eq:finite-boundary-shifts}
\end{align}
These relations follow by integrating eq.~\eqref{eq:coefficient-map} over the canonical disk and are consistent with the distribution-space boundary analysis in ref.~\cite{EbertTackmann2017}. They apply to the longitudinal distributions themselves, including their flavor indices.

The spectral construction below uses the parent boundary at $L_b=0$. When its fixed-order expansion is later integrated over a disk, eq.~\eqref{eq:finite-boundary-shifts} is generated automatically. Keeping the two boundaries distinct preserves the finite terms required for primed accuracy. Appendix~\ref{app:matching} gives the CS boundary translation and the ordered recurrence that reconstructs the logarithms through three loops.

\subsection{Cumulative scale setting and contact preservation}

Momentum-space cumulative scale setting is useful for fixed-order comparisons and alternative profile prescriptions~\cite{EbertTackmann2017}. Its definition must retain the origin contact. With $t=k_T^2$, let $\psi_\varphi(t)$ be the angular average of $\varphi(\bk)$ and let $C_p(t)$ be a profiled cumulative. We define
\begin{equation}
 \pair{\mathcal S_p A}{\varphi}
       =-\int_0^\infty\dd t\,C_p(t)\psi_\varphi'(t).
 \label{eq:weak-profile}
\end{equation}
For a locally integrable cumulative of polynomial growth, this is the radial lift of $\dd[\theta(t)C_p(t)]/\dd t$. Its disk action is $C_p(K^2)$. In particular, a constant cumulative $C_p(t)=c$ gives $c\deltaT$, including in the first bin. Logarithmically divergent origin cumulants are handled by the same weak action. The precise hypotheses, contact decomposition, and finite-order inheritance result are given in appendix~\ref{app:profiles}.

This cumulative operation and the spectral canonical construction are distinct realizations of scale setting. A cumulative profile generally does not commute with transverse convolution. The construction proved below instead solves the scalar canonical RG equations as a function of the spectrum of $T$ and then evaluates their complete operator. It thereby fixes the ordering of matching, running coupling, and evolution before acting on the fitted function.

\section{Unexpanded Collins--Soper evolution in momentum space}
\label{sec:unexpanded}

\subsection{Fixed-flavor running and an explicit infrared completion}

Work in massless QCD at fixed $n_f$, with $\beta_0>0$ and $\Gamma_0^i>0$. Retain
\begin{align}
 \beta_{[4]}(a)&=-2\sum_{r=0}^3\beta_r a^{r+2},&
 \Gamma_{i,[4]}(a)&=\sum_{r=0}^3\Gamma_r^i a^{r+1},\notag\\
 \gamma_{i,[3]}(a)&=\sum_{r=0}^2\gamma_r^i a^{r+1},&
 d_{i,[3]}(a)&=d_i^{(2)}a^2+d_i^{(3)}a^3.
 \label{eq:retained-functions}
\end{align}
The rapidity boundary has $d_i^{(1)}=0$ in this convention. The cusp, beta function, and rapidity coefficients are standard perturbative inputs~\cite{vanRitbergen1997,HKM2020,LiZhu2017}. At four loops the partonic color dependence is supplied from the full cusp result, including quartic invariants. It is not replaced by a universal quadratic-Casimir rescaling.

Select a fixed $\mu_0>0$ and $a(\mu_0)=a_0$ such that
\begin{equation}
 \beta_{[4]}(a)<0\quad\text{for }0<a\le a_0.
 \label{eq:AF-branch}
\end{equation}
The coupling is the unexpanded solution of its retained ordinary differential equation on $\nu\ge\mu_0$. All external virtuality scales lie in this interval. The chosen branch is asymptotically free and every scale integral below stays above $\mu_0$. Finite heavy-flavor thresholds and mass-dependent factors are separate extensions to this fixed-$n_f$ construction.

Let $s$ be the spectral value of $T=-\nabla_k^2$. It has units of inverse momentum squared and differs from both $k_T^2$ and $s_{\rm had}$. Define $s_0=b_0^2/\mu_0^2$ and the smooth step
\begin{equation}
\chi(u)=
 \begin{cases}
 0,&u\le0,\\
 \displaystyle\frac{e^{-1/u}}{e^{-1/u}+e^{-1/(1-u)}},&0<u<1,\\
 1,&u\ge1 .
 \end{cases}
 \label{eq:smooth-step}
\end{equation}
Set $c(y)=\chi(2y-1)$. Our canonical spectral scale is
\begin{equation}
 \mu_c(s)=\mu_0\exp\!\left[-\frac{1-c(s/s_0)}2\ln\frac{s}{s_0}\right],
       \qquad s>0.
 \label{eq:canonical-scale}
\end{equation}
It equals $b_0/\sqrt{s}$ for $s\le s_0/2$, is $\mu_0$ for $s\ge s_0$, and remains at least $\mu_0$ in between. It has no ultraviolet cap: $\mu_c(s)\to\infty$ as $s\to0$. The scale is independent of the hard scale, longitudinal fraction, hadron identity, and endpoint path, preserving the universality of the evolution kernel.

The completion in eq.~\eqref{eq:canonical-scale} defines the perturbative/long-distance separation in the infrared. It is exact canonical scale setting in the short-distance region. Two such completions give kernels whose difference is independent of $\mu$, since their $\mu$ derivatives coincide. Describing the same full TMD with another completion therefore entails a corresponding change of its long-distance CS contribution and boundary. This is the usual prescription dependence of the infrared separation expressed at the operator level~\cite{BillisMichelTackmann2025,CeruttiSimonelli2026}.

\subsection{RG-resummed CS kernel and endpoint solution}

Define
\begin{equation}
 D_i^c(s;\mu)
  =d_{i,[3]}\!\left(a(\mu_c(s))\right)
       +2\int_{\mu_c(s)}^\mu\frac{\dd\nu}{\nu}\,
                                      \Gamma_{i,[4]}(a(\nu)).
 \label{eq:D-resummed}
\end{equation}
Its derivative is exactly
\begin{equation}
 \partial_{\ln\mu}D_i^c(s;\mu)=2\Gamma_{i,[4]}(a(\mu))
 \label{eq:D-exact-RG}
\end{equation}
within the retained system. In the canonical region its expansion reproduces the one-, two-, and three-loop CS polynomial, including the term $2\beta_0d_i^{(2)}$ in the linear three-loop logarithm; see appendix~\ref{app:matching}. The logarithms have been resummed by integrating the RG equation before any exponential is formed. Unexpanded integration of the retained Sudakov evolution equations is the scalar perturbative operation used here~\cite{EbertSudakov2022}.

For $e_a=(\mu_a,\zeta_a)$ and $e_b=(\mu_b,\zeta_b)$, write
\begin{align}
 \rho_{ba}&=\ln\sqrt{\zeta_b/\zeta_a},\notag\\
 R_{i;ba}(\zeta_a)
  &=\int_{\mu_a}^{\mu_b}\frac{\dd\nu}{\nu}
      \left[\gamma_{i,[3]}(a(\nu))-
       \Gamma_{i,[4]}(a(\nu))\ln\frac{\zeta_a}{\nu^2}\right],\notag\\
 u_{i;ba}(s)&=\exp\!\left[R_{i;ba}(\zeta_a)-\rho_{ba}D_i^c(s;\mu_b)\right].
 \label{eq:endpoint-symbol}
\end{align}
Here the subscript $b$ labels the endpoint; it is distinct from the spectral canonical scale $\mu_c$. The momentum-space operators are
\begin{equation}
 \Uop_i(e_b;e_a)=u_{i;ba}(T),\qquad \Dop_i^c(\mu)=D_i^c(T;\mu),
 \label{eq:endpoint-operator}
\end{equation}
using the spectral calculus of the nonnegative self-adjoint Laplacian~\cite{Teschl2014}.

The order of operations is physically and mathematically important. At fixed $\mu$, the three-loop polynomial has a leading term $a^3\beta_0^2\Gamma_0^iL_b^3/3$. Exponentiating that polynomial literally at positive $\rho$ produces growth $\exp(c|\ln b_T|^3)$ near $b_T=0$, outside the regular tempered-multiplier domain. In contrast, the RG-resummed kernel in eq.~\eqref{eq:D-resummed} has only a logarithm-of-logarithm singularity there. The admitted endpoint solution follows from this improved asymptotic behavior.

\begin{theorem}[Admitted endpoint evolution]
\label{thm:endpoint}
Let $\tau>0$ have units of inverse momentum squared and define
\begin{equation}
\begin{aligned}
 w(s)&=1+\ln(1+\tau/s),\\
 \Dlog&=\bigcap_{m=0}^{\infty}\Dom w(T)^m\subset\Ltwo .
\end{aligned}
 \label{eq:domain}
\end{equation}
Under eqs.~\eqref{eq:AF-branch}--\eqref{eq:canonical-scale}, this is a dense common invariant domain containing $\Sch$. For finite endpoints the maps $\Uop_i(e_b;e_a)$ and their inverses are continuous in its seminorms. They obey
\begin{equation}
\begin{aligned}
 \Uop_i(e_c;e_b)\Uop_i(e_b;e_a)&=\Uop_i(e_c;e_a),\\
 \Uop_i(e_a;e_a)&=I,\\
 \partial_{\ln\sqrt{\zeta_b}}\Uop_i(e_b;e_a)
 &=-\Dop_i^c(\mu_b)\Uop_i(e_b;e_a),\\
 \partial_{\ln\mu_b}\Uop_i(e_b;e_a)
 &=\gamma_{F,i}^{[4,3]}(\mu_b,\zeta_b)\Uop_i(e_b;e_a).
\end{aligned}
 \label{eq:operator-evolution-laws}
\end{equation}
Here $\gamma_{F,i}^{[4,3]}(\mu,\zeta)=\gamma_{i,[3]}(a(\mu))-\Gamma_{i,[4]}(a(\mu))\ln(\zeta/\mu^2)$ is the retained TMD anomalous dimension. The identities are exact on $\Dlog$ for the retained RG system. The symbols are regular tempered multipliers, so their inverse Fourier images define distributional transverse evolution kernels.
\end{theorem}
\begin{proof}
On the chosen branch,
\begin{equation}
 a(\nu)=\frac{1}{2\beta_0\ln(\nu/\nu_*)}
       +\Order{\frac{\ln\ln(\nu/\nu_*)}{\ln^2(\nu/\nu_*)}}
 \label{eq:UV-running}
\end{equation}
as $\nu\to\infty$, for a fixed positive reference $\nu_*$. Therefore
\begin{equation}
 D_i^c(s;\mu)
       =-\frac{\Gamma_0^i}{\beta_0}
            \ln\ln\frac{\mu_c(s)}{\nu_*}+O(1),\qquad s\to0.
 \label{eq:D-asymptotic}
\end{equation}
The canonical boundary tends to zero. On compact sets of endpoints there are constants $C,p<\infty$ such that $|u_{i;ba}|+|u_{i;ba}^{-1}|\le Cw^p$. At $s\ge s_0$ the symbols are constant. Powers of logarithms are locally integrable in the two-dimensional spectral measure $\pi\dd s$, so the symbols are tempered. Moreover,
\begin{equation}
 \norm{w(T)^m\Uop_i f}_2\le C\norm{w(T)^{m+\lceil p\rceil}f}_2.
 \label{eq:domain-bound}
\end{equation}
This gives invariance and continuity. Schwartz Fourier images are bounded near zero and rapidly decreasing at infinity, which proves inclusion and density. The maximal spectral multiplier is a closed operator, and the common domain is its invariant restriction.

Differentiating eq.~\eqref{eq:endpoint-symbol} using eq.~\eqref{eq:D-exact-RG} gives the two endpoint derivatives. The $\mu_b$ derivative combines $\zeta_a$ in $R$ with $2\rho_{ba}$ to give $\zeta_b$. Integrating the same identity establishes equality of rectangular paths and the composition law. Differentiations are controlled by higher seminorms in eq.~\eqref{eq:domain-bound}.
\end{proof}

At one-loop running with $d=0$, the result is especially transparent:
\begin{equation}
 D_i^c(s;\mu)=\frac{\Gamma_0^i}{\beta_0}
                    \ln\frac{a(\mu_c(s))}{a(\mu)}.
 \label{eq:D-one-running}
\end{equation}
The endpoint factor has at most logarithmic-power growth at the ultraviolet spectral endpoint. This also shows why the common domain is more appropriate than demanding boundedness of every endpoint ratio on all of $L^2$.

\subsection{Bounded canonical legs}

A canonical input can be evolved directly with
\begin{equation}
\begin{aligned}
 \ln v_i(s;\mu,\zeta)
 &=-d_{i,[3]}(a(\mu_c))\ln\frac{\sqrt\zeta}{\mu_c}\\
 &\quad+\int_{\mu_c}^{\mu}\frac{\dd\nu}{\nu}
                      \gamma_{F,i}^{[4,3]}(\nu,\zeta).
\end{aligned}
 \label{eq:canonical-symbol}
\end{equation}
It solves evolution from the spectral point $(\mu_c,\mu_c^2)$ and satisfies
\begin{equation}
 u_{i;ba}(s)=\frac{v_i(s;e_b)}{v_i(s;e_a)}.
 \label{eq:ratio}
\end{equation}

\begin{theorem}[Bounded canonical evolution]
\label{thm:canonical}
For a fixed finite endpoint, $v_i$ is positive for $s>0$, extends continuously by $v_i(0)=0$, is bounded, and is constant for $s\ge s_0$. More precisely,
\begin{equation}
 \ln v_i(s;\mu,\zeta)
  =-\frac{\Gamma_0^i}{\beta_0}\ln\frac{\mu_c(s)}{\nu_*}
        +\Order{\left[\ln\ln\frac{\mu_c(s)}{\nu_*}\right]^2}.
 \label{eq:v-asymptotic}
\end{equation}
For every $0<\epsilon<p_i=\Gamma_0^i/(2\beta_0)$,
$v_i(s)=O((s/s_0)^{p_i-\epsilon})$ near zero. The operator $v_i(T;e)$ is bounded on $L^2$, has dense range, and maps $L^2$ into $\Dlog$.
\end{theorem}
\begin{proof}
The large-$\nu$ cusp term $2\Gamma_{i,[4]}(a(\nu))\ln(\nu/\nu_*)$ tends to $\Gamma_0^i/\beta_0$. Integrating it from $\mu_c$ down to the fixed endpoint gives the leading negative term in eq.~\eqref{eq:v-asymptotic}. The subleading running terms contribute at most squared logarithms of $\ln\mu_c$; the noncusp integral grows only logarithmically in $\ln\mu_c$. Since $d$ starts at $a^2$, its boundary term is $O(1/\ln\mu_c)$. The canonical relation $\mu_c=b_0/\sqrt{s}$ then gives the power estimate. Boundedness follows by continuity on the remaining finite interval and constancy in the infrared. Multiplication by any power of $w$ preserves boundedness, proving the range inclusion. Positivity almost everywhere gives a trivial kernel and dense range.
\end{proof}

For longitudinal matching define
\begin{equation}
\begin{aligned}
 B_{i/h}^{[n_B]}(x;s)
 &=\sum_j[C_{i\leftarrow j}^{\rm can,[n_B]}(a_c)
                       \ox f_{j/h}(\mu_c)](x),\\
 M_{i/h}(x,s;e)&=v_i(s;e)B_{i/h}^{[n_B]}(x;s).
\end{aligned}
 \label{eq:canonical-matching}
\end{equation}
Here $\mu_c=\mu_c(s)$, $a_c=a(\mu_c(s))$, and $C^{\rm can,[n_B]}$ is the parent boundary through order $a^{n_B}$ at $L_b=\ell_\zeta=0$. The operator-domain argument applies at any specified finite $n_B$ satisfying the following bounds. In the validated physical implementation $n_B=1$, and the assembled hard/OPE products are truncated strictly at NLO. Assume these longitudinal actions exist, are continuous for finite $\mu_c$, and grow at most polynomially in $\ln(\mu_c/\nu_*)$, with analogous derivative bounds. Fixed Mellin moments away from splitting-kernel singularities satisfy a sufficient version of this condition: Gronwall's inequality applied to finite-order DGLAP evolution gives at most powers of $\ln\mu_c$. Uniform $x$-space use requires the corresponding bounds on the selected inputs. We use compact longitudinal regions satisfying them.

Under these conditions $M_{i/h}$ is bounded, tends to zero at $s=0$, and is constant for $s\ge s_0$. Signed matching coefficients are allowed. The ultraviolet power in theorem~\ref{thm:canonical} dominates the assumed longitudinal growth. Thus the complete leg can act directly on a canonical input in $L^2$ using eq.~\eqref{eq:central-leg}.

The zero endpoint $M(0)=0$ characterizes this ultraviolet-uncapped singular-region solution. If a nonzero output is integrable, its transverse integral equals its Fourier image at zero and therefore vanishes. Such an output cannot be everywhere nonnegative. Positivity of a localized input is consequently not a positivity theorem for the complete singular contribution. Signed large-$q_T$ bins are retained in the numerical stress tests. An inclusive cross section is obtained from its matched measurement, not from interpreting that endpoint as a probability normalization. The separate roles of input normalization, regulated TMD moments, and full-spectrum matching are detailed in sections~\ref{sec:extraction} and \ref{sec:observables}.

\section{Constructive evaluation by momentum-space resolvents}
\label{sec:evaluation}

\subsection{A momentum-space realization of the functional calculus}

An operator definition becomes useful for an extraction when its action can be approximated without transforming the fitted function. Introduce
\begin{equation}
 R_\tau=\tau(\tau+T)^{-1},\qquad T=-\nabla_k^2,
 \label{eq:resolvent}
\end{equation}
where $\tau>0$ has units of inverse momentum squared, as does the spectrum $s$ of $T$. The resolvent is a bounded positive self-adjoint contraction. It acts by a transverse convolution,
\begin{equation}
\begin{aligned}
 (R_\tau f)(\bk)&=\int\dd^2\bp\,r_\tau(\bk-\bp)f(\bp),\\
 r_\tau(\bk)&=\frac{\tau}{2\pi}K_0(\sqrt\tau\,k_T).
\end{aligned}
 \label{eq:resolvent-kernel}
\end{equation}
The kernel follows by integrating the two-dimensional heat kernel against $\tau e^{-\tau u}\dd u$ and satisfies $\int\dd^2\bk\,r_\tau=1$. Its logarithmic singularity is locally integrable. Equivalently, applying $R_\tau$ is the elliptic solve
\begin{equation}
 (\tau-\nabla_k^2)\,g=\tau f.
 \label{eq:elliptic-solve}
\end{equation}
The modified Bessel function in eq.~\eqref{eq:resolvent-kernel} is an ordinary kernel function; neither realization is a numerical Bessel transformation of $f$.

For the admitted canonical symbol $M(s)$ of eq.~\eqref{eq:canonical-matching}, suppress the parton, hadron, and longitudinal arguments and put
\begin{equation}
\begin{aligned}
 M_\infty&=\lim_{s\to\infty}M(s),\qquad r=\frac{\tau}{\tau+s},\\
 m(r)&=M\!\left(\tau\frac{1-r}{r}\right),\\
 h(r)&=\frac{m(r)-M_\infty}{r}\quad(r>0),\qquad h(0)=0 .
\end{aligned}
 \label{eq:compactification}
\end{equation}
The infrared completion makes $h$ identically zero near $r=0$. The ultraviolet limit $M(0)=0$ gives $h(1)=-M_\infty$. Thus $h\in C[0,1]$. Define the Bernstein polynomial
\begin{equation}
 B_N[h](r)=\sum_{j=0}^N h(j/N)\binom Nj r^j(1-r)^{N-j}
 \label{eq:bernstein}
\end{equation}
and the momentum-space approximant
\begin{equation}
 \mathbb M_N=M_\infty I+R_\tau B_N[h](R_\tau).
 \label{eq:evaluator}
\end{equation}
The scalar samples $h(j/N)$ require ordinary one-dimensional running-coupling and matching calculations. They encode the same transverse resolution as the parent spectral variable. The fitted input itself remains a function of $\bk$: eq.~\eqref{eq:evaluator} acts on it through resolvents and linear combinations.

\begin{theorem}[Contact-preserving momentum-space evaluator]
\label{thm:evaluator}
Let $M$ obey the canonical endpoint properties and longitudinal growth assumptions of section~\ref{sec:unexpanded}, and define $\epsilon_N=\norm{B_N[h]-h}_\infty$. Then
\begin{align}
 \norm{\mathbb M_N-M(T)}_{2\to2}&\le\epsilon_N\longrightarrow0,\notag\\
 M(T)\deltaT&=M_\infty\deltaT+g,\qquad g\in\Ltwo,\notag\\
 \norm{(\mathbb M_N-M(T))\deltaT}_2
       &\le\sqrt{\frac{\tau}{4\pi}}\,\epsilon_N.
 \label{eq:evaluator-bounds}
\end{align}
The contact coefficient is exact at every approximation degree. If $\mathcal B$ has finite transverse area and a specified origin-inclusion convention, then
\begin{equation}
 \left|\int_{\mathcal B}\dd^2\bk\,
       (\mathbb M_N-M(T))\deltaT\right|
       \le\sqrt{|\mathcal B|}\sqrt{\frac{\tau}{4\pi}}\,\epsilon_N.
 \label{eq:contact-bin-bound}
\end{equation}
If $h$ is H\"older continuous with exponent $0<\nu\le1$ and constant $H_\nu$, a sufficient rate is
\begin{equation}
 \epsilon_N\le H_\nu(4N)^{-\nu/2}.
 \label{eq:holder-rate}
\end{equation}
\end{theorem}
\begin{proof}
The spectral multiplier error is $r(B_N[h](r)-h(r))$. Since $0\le r\le1$, the spectral theorem gives the operator bound. Uniform Bernstein convergence follows from continuity on $[0,1]$. The exact $r=0$ value is $M_\infty$, so the contact coefficient is preserved.

For the contact input, $M-M_\infty$ is bounded and supported at finite $s$. Its inverse Fourier image is therefore in $L^2$. The leading resolvent also makes the approximate remainder square integrable. For proof only, Plancherel with eq.~\eqref{eq:fourier} gives
\begin{equation}
 \norm{g_N-g}_2^2
 \le\frac{\epsilon_N^2}{(2\pi)^2}
       \int\dd^2\bb\left(\frac{\tau}{\tau+b_T^2}\right)^2
 =\frac{\tau}{4\pi}\epsilon_N^2.
 \label{eq:contact-plancherel}
\end{equation}
Cauchy--Schwarz gives the bin bound. Finally write $B_N[h](r)=\mathbb E[h(J/N)]$ for a binomial random variable $J$ with parameters $(N,r)$. The H\"older condition and Jensen's inequality imply
$|B_N[h](r)-h(r)|\le H_\nu[r(1-r)/N]^{\nu/2}$, proving eq.~\eqref{eq:holder-rate}.
\end{proof}

For the canonical RG symbol, the differentiated ultraviolet estimates give H\"older continuity for every $0<\nu<\min(1,p_i)$, where $p_i=\Gamma_0^i/(2\beta_0)$. The same conclusion applies to longitudinal inputs satisfying the stated derivative bounds. This rate provides a constructive convergence result. A $C^2$ Bernstein estimate is available for smoother symbols, but ultraviolet differentiability must be checked before it is used. The exponent $\nu$ may be small, so the existence of a convergent evaluator and its eventual computational efficiency are distinct questions.

\subsection{A practical recurrence and observable error bounds}

The polynomial can be applied without forming monomial coefficients. For an input $f$, initialize
\begin{equation}
 q_j^{(0)}=h(j/N)f,\qquad j=0,\ldots,N,
 \label{eq:decasteljau-initial}
\end{equation}
and iterate
\begin{equation}
 q_j^{(\ell+1)}=(I-R_\tau)q_j^{(\ell)}+R_\tau q_{j+1}^{(\ell)},
 \qquad j=0,\ldots,N-\ell-1.
 \label{eq:decasteljau}
\end{equation}
The result is $\mathbb M_N f=M_\infty f+R_\tau q_0^{(N)}$. This operator de Casteljau recurrence consists of repeated applications of the same momentum-space elliptic solver. It also makes explicit which operations can be reused among different fitted inputs. The continuum theorem concerns the exact $R_\tau$; a discretized solver requires its own finite-volume, grid, and solver-tolerance study.

For ordinary canonical inputs, let $F_A,F_B\in L^2$ and let their approximations have errors bounded by $e_A,e_B$ in that norm. Their simple transverse-convolution bin then obeys
\begin{align}
 &\left|\int_{\mathcal B}\dd^2\bq\,
       (F_{A,N}\ot F_{B,N}-F_A\ot F_B)(\bq)\right|\notag\\
 &\qquad\le |\mathcal B|\left(e_A\norm{F_B}_2+
                      e_B\norm{F_A}_2+e_Ae_B\right).
 \label{eq:two-leg-bound}
\end{align}
This follows by expanding the difference and applying Cauchy--Schwarz at each $\bq$. Flavor sums and absolute hard factors can be included term by term. More general measurements inherit bounds from their own continuity properties.

For perturbative contact legs, it is convenient to apply theorem~\ref{thm:evaluator} to the complete product symbol. Each canonical symbol is constant for sufficiently large $s$, so the product again has a constant plus a compactly supported $L^2$ spectral remainder. This treats the contact state through its admitted multiplier product, rather than applying an $L^2$ input theorem to $\deltaT$.

Arbitrary endpoint inputs in $\Dlog$ can also be evaluated. Multiply $u_{i;ba}$ by a smooth spectral cutoff that removes $s\le\varepsilon$, apply the bounded evaluator, and remove the cutoff in the strong topology of the common domain. Appendix~\ref{app:operator} gives the explicit bound. The canonical construction is especially convenient for fitting because it applies a bounded operator directly, without this additional limit.

\subsection{Endpoint-preserving Chebyshev evaluation}
\label{sec:chebyshev}

The Bernstein construction provides a convergence proof and a direct
implementation. For the high-resolution benchmarks we instead interpolate the
compactified function $m(r)$ at Chebyshev--Lobatto nodes, including both
$r=0$ and $r=1$. Let $X=2R_\tau-I$ and
\begin{equation}
 p_N(r)=\sum_{j=0}^N c_j T_j(2r-1).
 \label{eq:chebyshev}
\end{equation}
Endpoint interpolation enforces $p_N(0)=M_\infty$ and $p_N(1)=0$.
The forward polynomial recurrence is
\begin{align}
 y_0&=f,\qquad y_1=Xf,\notag\\
 y_{j+1}&=2Xy_j-y_{j-1},\qquad
 p_N(R_\tau)f=\sum_{j=0}^N c_jy_j .
 \label{eq:cheb-action}
\end{align}
Each application of $X$ uses one momentum-space elliptic solve and vector
operations. An equivalent Clenshaw recurrence needs three input-shaped work
arrays. Neither path transforms the input. For many physical components with
the same input, we cache each bin's action on $y_j$ and contract those moments
with the component-specific coefficients. This reuses the expensive solves
without changing the multiplier being approximated.

On an azimuthally symmetric grid, the Laplacian is represented in the
$2\pi k\,\dd k$ measure. A regular zero-flux condition is imposed at the origin
and a homogeneous Dirichlet condition at the outer radius. Similarity by the
square roots of the cell weights produces a symmetric discrete operator.
The exact continuum and finite-grid operators are distinct: increasing the
polynomial degree at a fixed grid does not remove radial discretization or
finite-volume error.

Theorem~\ref{thm:evaluator} proves the Bernstein sequence's convergence and its
weighted contact bound. It does not automatically certify the error or rate of
the Chebyshev sequence. The latter preserves the endpoints and is tested against
separate spectral/Clenshaw implementations, a continuum reference, and independent
degree, spacing and radius refinements. The reported numerical budgets concern
those observed differences; they are not sampled estimates relabeled as a
rigorous global operator norm.

\section{Logarithmic accuracy and perturbative matching}
\label{sec:accuracy}

\subsection{The retained canonical exponent}

The accuracy statement refers to the perturbative canonical region
$0<s\le s_0/2$, with $\mu_c(s)\le Q$. Define
\begin{equation}
 a_Q=a(Q),\qquad L=\ln\frac{Q}{\mu_c(s)},\qquad a_-(t)=a(Qe^{-t}).
 \label{eq:log-variables}
\end{equation}
At the symmetric endpoint $(\mu,\zeta)=(Q,Q^2)$, the exponent of the canonical evolution factor is
\begin{equation}
\begin{aligned}
 E_{i,[4,3,3,4]}(L,a_Q)
 &=\int_0^L\dd t\,\gamma_{i,[3]}(a_-(t))\\
 &\quad-2\int_0^L\dd t\,t\Gamma_{i,[4]}(a_-(t))\\
 &\quad-L\,d_{i,[3]}(a_-(L)).
\end{aligned}
 \label{eq:resummed-exponent}
\end{equation}
The index records the retained cusp, noncusp, CS-boundary, and beta-function orders. Equation~\eqref{eq:resummed-exponent} defines the numerical exponent without expanding its running coupling. The organization is the canonical CSS2 RG solution~\cite{CollinsRogers2017,EbertSudakov2022}; the operator calculus establishes how to evaluate it on momentum-space inputs.

\begin{theorem}[All-order logarithmic towers]
\label{thm:accuracy}
With the retained functions in eq.~\eqref{eq:retained-functions}, the formal expansion of eq.~\eqref{eq:resummed-exponent} contains the complete CSS2 exponent towers
\begin{equation}
 a_Q^nL^{n+1},\qquad a_Q^nL^n,\qquad
 a_Q^nL^{n-1},\qquad a_Q^nL^{n-2},
 \label{eq:retained-towers}
\end{equation}
wherever the indicated powers occur. The first omitted perturbative inputs change the exponent only by
\begin{equation}
 \Delta E_i\in\sum_{n\ge4}a_Q^n\mathbb C[L]_{\le n-3}.
 \label{eq:omitted-towers}
\end{equation}
The evolution therefore has $N^3\mathrm{LL}$ accuracy in the declared convention. This theorem concerns the retained exponent; finite hard and matching boundaries are separate ingredients.
\end{theorem}
\begin{proof}
Write the downward-running solution as $a_-(t)=\sum_{n\ge1}a_Q^nA_n(t)$, with $A_1=1$. Substitution in
\begin{equation}
 \frac{\dd a_-}{\dd t}=2\sum_{r\ge0}\beta_r a_-^{r+2}
 \label{eq:downward-ODE}
\end{equation}
proves by induction that $\deg A_n\le n-1$. More generally, the coefficient of $a_Q^n$ in $a_-^j$ has degree at most $n-j$.

The first omitted cusp coefficient multiplies $a_-^5$. The extra factor of $t$ and integration in eq.~\eqref{eq:resummed-exponent} give degree at most $n-3$ at order $n$. The first omitted noncusp coefficient starts at $a_-^4$, whose single integral has the same maximum degree. The omitted CS boundary is $a_-^4(L)$ multiplied by $L$, again giving $n-3$.

The first omitted beta coefficient multiplies $a_-^6$. Its first effect on the running solution is $O(a_Q^6t)$; induction in the varied RG equation bounds its order-$n$ coefficient by degree $n-5$. Inserting that correction into the leading cusp integral gives at most degree $n-3$. Insertions in higher cusp terms, the noncusp integral, and the CS boundary are less enhanced. Products generated by the recurrence satisfy the same degree inequalities. This establishes eq.~\eqref{eq:omitted-towers} at every order, not just through $a_Q^3$.

The retained coefficients are those of CSS2 because the equations, canonical boundaries, and integral solution coincide. The continuum functional calculus represents the same complete symbol and contributes no additional perturbative tower. Finite-order witnesses can check this induction at selected orders, but the induction itself supplies the all-order statement.
\end{proof}

As a useful closed limit, one-loop running and the leading cusp give
\begin{equation}
 E_i^{\rm LL}
 =\frac{\Gamma_0^i}{2\beta_0^2a_Q}
       [\ln(1-\lambda)+\lambda],
 \qquad \lambda=2\beta_0a_QL<1.
 \label{eq:LL-exponent}
\end{equation}
This expression exhibits the infinite leading-logarithmic tower explicitly. Its sign follows the Sudakov suppression of the canonical solution. A fixed-order comparison tests coefficients of this tower; the RG argument establishes its all-order organization.

\subsection{Uniform accuracy on a specified perturbative window}

The formal logarithmic theorem can be strengthened when the comparison RG functions have controlled remainders. This distinguishes perturbative tower counting from a convergence assumption about the full QCD series.

\begin{proposition}[Uniform resummed remainder]
\label{prop:uniform}
Suppose the comparison functions satisfy
\begin{equation}
\begin{aligned}
 \Delta\beta&=O(a^6),&\Delta\Gamma&=O(a^5),\\
 \Delta\gamma&=O(a^4),&\Delta d&=O(a^4).
\end{aligned}
 \label{eq:remainder-inputs}
\end{equation}
with the corresponding local derivative bounds. Consider a family of intervals with $0\le a_QL\le\lambda_*$ below the first RG singularity and $a_-(t)\le C a_Q$ uniformly. The exponent difference is $O(a_Q^3)$ uniformly on this window. If the finite hard/matching boundaries are retained through $a^3$, their boundary remainder is $O(a_Q^4)$.
\end{proposition}
\begin{proof}
Variation of the running-coupling ODE and Gronwall's inequality give
$\Delta a_-=O(a_Q^6L)=O(a_Q^5)$. The omitted cusp term and the change in the running coupling enter a double-logarithmic integral, yielding $O(a_Q^5L^2)=O(a_Q^3)$. The noncusp integral and CS boundary contribute $O(a_Q^4L)=O(a_Q^3)$. A finite fourth-order boundary carries no further large logarithm and is $O(a_Q^4)$. The constants are uniform on the selected compact resummation window.
\end{proof}

The resulting exponent organization is
\begin{equation}
\begin{aligned}
 E_i&=\frac{1}{a_Q}g_{1,i}(a_QL)+g_{2,i}(a_QL)\\
 &\quad+a_Qg_{3,i}(a_QL)+a_Q^2g_{4,i}(a_QL)+O(a_Q^3).
\end{aligned}
 \label{eq:g-functions}
\end{equation}
under these hypotheses. The infrared completion supplies the continuation outside the perturbative canonical window. Its fitted long-distance counterpart is specified separately; the uniform estimate concerns the stated perturbative region and does not assign a perturbative error to unknown infrared structure.

\subsection{Higher-order assembly and the implemented accuracy}

The distributional map is not restricted to one-loop boundaries. Given
scheme-consistent physical coefficients, it projects each retained order into
the same transverse algebra. A first omitted hard or canonical matching
boundary after third order starts at $a^4$; reexpression through the running
coupling generates degree at most $n-4$ at order $a_Q^n$. Collinear evolution
retains its ordered flavor products. This is the formal organization required
for an $\Nthree$ singular prediction, with the additional ingredients in
Table~\ref{tab:accuracy}. It is a specification for extending the implementation,
not a claim that the corresponding physical three-loop boundary payload has
been implemented or numerically checked here.

\begin{table*}[t]
\caption{The general high-order assembly and the accuracy actually instantiated
in the numerical study. The present label is $N^3\mathrm{LL}$ evolution with
strict-NLO hard/OPE boundaries. Neither a PDF-set order nor a formal coefficient
recurrence promotes that realization to a primed-order spectrum.}
\label{tab:accuracy}
\centering
\begin{tabular}{@{}p{0.25\textwidth}p{0.31\textwidth}p{0.36\textwidth}@{}}
\toprule
\raggedright Ingredient & \raggedright Formal $\Nthree$ singular prescription & \raggedright Present physical realization\tabularnewline
\midrule
\raggedright Cusp and beta function & \raggedright Four loops & \raggedright Four-loop sourced quark cusp and fixed-$n_f$ running\tabularnewline
\raggedright Noncusp anomalous dimension & \raggedright Three loops & \raggedright Three-loop quark TMD noncusp input\tabularnewline
\raggedright Canonical CS boundary & \raggedright Through $a^3$ & \raggedright $d^{(1)}=0$, physical $d^{(2)},d^{(3)}$\tabularnewline
\raggedright Hard and TMD matching boundaries & \raggedright Through $a^3$ in a common scheme & \raggedright Strict-NLO hard and OPE insertions\tabularnewline
\raggedright Collinear ingredients & \raggedright Consistent PDF/evolution and flavor conventions & \raggedright Pinned NNPDF4.0 NNLO table; separate fixed-$n_f=5$ running\tabularnewline
\raggedright Product expansion & \raggedright Every $r+s+t=n$ hard/two-leg partition & \raggedright Six linear Born/hard/$qq$/$qg$ terms; no incomplete products\tabularnewline
\raggedright Nonsingular remainder & \raggedright Separate fixed-order and measurement specification & \raggedright Full $W+Y$ implementation deferred\tabularnewline
\bottomrule
\end{tabular}
\end{table*}

Published unpolarized matching functions through N$^3$LO
\cite{EbertMistlbergerVita2020,EbertMistlbergerVita2021FF,LuoYangZhuZhu2021}
and the NNLO reference \cite{EchevarriaScimemiVladimirov2016} supply possible
finite inputs. The rapidity, cusp, running, splitting and hard functions have
their own source and convention checks
\cite{LiZhu2017,HKM2020,vanRitbergen1997,MochVermaserenVogt2004,
VogtMochVermaseren2004,GehrmannGloverHuberIkizlerliStuderus2010}.
A finite scheme translation acts on the complete factorized product; a
transverse-dependent conversion cannot be absorbed into a transverse-independent
hard function alone \cite{CollinsRogers2017}.

At fixed common scale, the order-$n$ singular coefficient is
\begin{equation}
 W_n=\sum_{r+s+t=n} H_r\,[A_s\ot B_t].
 \label{eq:product-partitions}
\end{equation}
Flavor sums, longitudinal convolutions and the measurement act on this full
expression. Checking only the genuine $n$-loop term would omit required
lower-order products. The parent-to-disk translations in
Eqs.~\eqref{eq:finite-boundary-shifts} and \eqref{eq:CS-disk-shift} likewise
belong to a declared boundary convention; applying both conventions at once
would double count finite contacts.

Polynomial approximation is a separate numerical expansion. Its error is
controlled through the admitted symbol and input class, and numerical
tolerances are allocated independently of the perturbative label. An existence
bound permits a sufficiently accurate polynomial to be chosen; practical
degree, grid, RG and longitudinal errors still require separate measurements.
Absolute or peak-normalized errors remain meaningful in cancellation-sensitive
bins where a uniform relative criterion does not.

\section{A differentiable interface for hadronic transverse structure}
\label{sec:extraction}

\subsection{Perturbative forward operator and long-distance input}

The canonical construction separates the operator fixed by the chosen perturbative inputs and infrared prescription from the hadronic function inferred from data. Factorization fixes the evolution and short-distance structure, while an infrared completion specifies how that perturbative information is continued into its long-distance model domain. For an admitted localized input $S_{i/h,\theta}(x,\bk)$, the TMD is
\begin{equation}
 F^c_{i/h,\theta}(x;\mu,\zeta)
      =M_{i/h}(x,T;\mu,\zeta)\,S_{i/h,\theta}(x).
 \label{eq:fitted-TMD}
\end{equation}
A sufficient class has radial $S\in L^1\cap L^2$, unit integral, and controlled transverse moments. The same canonical input is used at every external endpoint. Equations~\eqref{eq:ratio} and \eqref{eq:canonical-matching} then give the universal endpoint evolution of the fitted TMD. Its dependence on the hard scale is generated by the perturbative operator and any separately specified universal nonperturbative CS factor.

This provides the perturbative extension of the momentum-space inverse strategy motivating the paper. The normalized profile of ref.~\cite{FernandoKellerMomentum2026} is an example of how a flexible function can be inferred through a transverse convolution. Here the learned function occupies the long-distance input position in eq.~\eqref{eq:fitted-TMD}, while the matching tail and retained logarithmic scale dependence enter through $M_{i/h}$. Shared flavor profiles, separate quark and antiquark profiles, or conditioning on flavor are choices to be tested against the available experimental sensitivity. The perturbative flavor matrices are retained in each case.

Normalization of $S$ fixes a property of that model component. It leaves the renormalized TMD, with its perturbative distributions and scale dependence, subject to the regulated integral relations in section~\ref{sec:observables}. Positivity may likewise be imposed on a broadening input when appropriate for the model class; physical predictions are formed from the complete renormalized operator.

Alternative nonperturbative interfaces include an additive core with an explicit overlap subtraction, or a separate contribution to the CS kernel. These are distinct parameterization classes with their own OPE and evolution constraints. The convolutional class in eq.~\eqref{eq:fitted-TMD} gives a constructive sufficient domain for the operator and differentiability results presented here.

\subsection{A compatible neural realization}

To illustrate how the domain conditions can be enforced, consider the following compatible realization of the input class. Let $\omega(k_T^2)$ be a fixed, radial, smooth, nonnegative envelope of compact support, and let $n_{i/h,\theta}(x,k_T^2)$ be a smooth neural output that is bounded on that support for the parameter range under consideration. Define
\begin{align}
 S_{i/h,\theta}(x,\bk)
  &=\frac{\omega(k_T^2)e^{n_{i/h,\theta}(x,k_T^2)}}{Z_{i/h,\theta}(x)},\notag\\
 Z_{i/h,\theta}(x)
  &=2\pi\int_0^\infty\dd k_T\,k_T\,
          \omega(k_T^2)e^{n_{i/h,\theta}(x,k_T^2)}.
 \label{eq:neural-input}
\end{align}
Assume $Z$ stays bounded away from zero. Smooth activations and dependence on $k_T^2$ provide regularity at the origin, and the envelope supplies localization and finite moments. This is one proposed extraction realization of the admitted class, rather than a restriction on the theorem or a fitted result. A spline or analytic model can replace $n_\theta$ without changing the perturbative operator.

The envelope specifies a long-distance model scale whose stability must be assessed in an extraction. It does not truncate the perturbative TMD tail: the action of $M(T)$ generates the known short-distance behavior outside the support of the localized input. For a compact radial smearing and a sufficiently regular ordinary perturbative tail, Taylor expansion gives
\begin{equation}
\begin{aligned}
 [S\ot F^{\pert}](\bk)
 &=F^{\pert}(\bk)+\frac{m_2}{4}\nabla_k^2F^{\pert}(\bk)+R_4(\bk),\\
 m_2&=\int\dd^2\bp\,p_T^2S(\bp).
\end{aligned}
 \label{eq:tail-preservation}
\end{equation}
away from the contact region. Appendix~\ref{app:inputs} gives the assumptions and remainder bound. This relates input localization to preservation of the leading perturbative tail, in line with the consistency requirements emphasized in refs.~\cite{GonzalezHernandezRainaldiRogers2023,AslanBoglioneGHRRainaldiRogersSimonelli2024}.

For a parameter component $\theta_a$, the normalized input has derivative
\begin{equation}
 \partial_{\theta_a}S_\theta
 =S_\theta\left(\partial_{\theta_a}n_\theta
       -\int\dd^2\bk\,S_\theta\partial_{\theta_a}n_\theta\right),
 \label{eq:normalized-gradient}
\end{equation}
with $x,i,h$ held fixed. This formula makes the radial measure and normalization constraint explicit. Other admissible network constructions can be used if their derivatives have the necessary $L^2$ bounds.

\begin{corollary}[Differentiation of the canonical forward map]
\label{cor:derivatives}
Assume $\theta\mapsto S_\theta$ is continuously differentiable as a map into $L^2$, and fix the perturbative input and scale prescription. Then
\begin{equation}
\begin{aligned}
 \partial_{\theta_a}F_\theta&=M(T)\partial_{\theta_a}S_\theta,\\
 \norm{\partial_{\theta_a}F_{\theta,N}
                -\partial_{\theta_a}F_\theta}_2
 &\le\epsilon_N\norm{\partial_{\theta_a}S_\theta}_2.
\end{aligned}
 \label{eq:forward-gradient}
\end{equation}
The two-leg bin prediction is differentiable, with its derivative obtained by differentiating either leg in turn.
\end{corollary}
\begin{proof}
Bounded linear operators preserve $L^2$ differentiability. Apply this fact and theorem~\ref{thm:evaluator} to the parameter difference quotient and its limit. The bilinear convolution map from $L^2\times L^2$ to a finite-area bin is continuous by eq.~\eqref{eq:two-leg-bound}, so the product rule applies.
\end{proof}

This is an application of the operator bounds to the extraction problem. It supplies a controlled continuum derivative, rather than relying solely on automatic differentiation of a finite grid. On a numerical grid, the same resolvent applications can be used in the forward and derivative recurrences. For a real scalar symbol the operator is self-adjoint, which also simplifies an adjoint implementation. If parameters modify the CS kernel or the scale completion itself, their derivatives additionally act on $M$; they are not covered by holding $M$ fixed in eq.~\eqref{eq:forward-gradient}.

\subsection{Scale-dependent long-distance broadening}

A nonperturbative CS component may be introduced relative to a fixed rapidity reference $\zeta_r$:
\begin{equation}
 F_{i/h}(\zeta)
 =\exp\!\left[-\ln\sqrt{\zeta/\zeta_r}\,D_i^{\NP}(T)\right]
      F^c_{i/h}(\zeta),
 \label{eq:NP-CS}
\end{equation}
with a separately specified product domain. The kernel $D_i^{\NP}$ is independent of $\mu$ in an RG-consistent separation, common to the appropriate partonic universality class, and suppressed at short distance. Hadron and $x$ dependence belong in the boundary input rather than an arbitrary process-dependent CS kernel.

A simple constructive example is $D_i^{\NP}(T)=g_iT$ with $g_i\ge0$. For $\rho=\ln\sqrt{\zeta/\zeta_r}\ge0$, the additional factor is the momentum-space heat semigroup,
\begin{equation}
 U_i^{\NP}(\bk;\rho)=\frac{1}{4\pi g_i\rho}
              e^{-k_T^2/(4g_i\rho)},
 \label{eq:NP-heat}
\end{equation}
with the delta limit at $g_i\rho=0$. It has unit normalization and second moment $4g_i\rho$. Its Fourier image is $e^{-\rho g_i b_T^2}$, showing explicitly how a short-distance-suppressed CS term produces transverse broadening. A backward interval requires a smaller domain, as expected for backward heat evolution. This example fixes the meaning of the module interface; a physical kernel is constrained by data or other nonperturbative information.

Changes in the perturbative infrared completion alter the separation between $D_i^c$ and $D_i^{\NP}$. Since the difference of two completed perturbative kernels is $\mu$ independent, the total kernel can be held fixed by an opposite change of the nonperturbative contribution, accompanied by the matching boundary adjustment. This makes prescription dependence explicit in a joint analysis of boundary structure and evolution.

\subsection{Information extracted from the data}

The resulting fit is an inverse problem for $S_\theta$ and, where desired, parameters in an admitted $D^{\NP}$. For a set of bins $r$, write the forward predictions as $\sigma_r(\theta)$ and a covariance-weighted objective as
\begin{equation}
 \mathcal J(\theta,\boldsymbol\nu)
 =\bigl[\boldsymbol\sigma(\theta,\boldsymbol\nu)-\boldsymbol d\bigr]^T
       V^{-1}
       \bigl[\boldsymbol\sigma(\theta,\boldsymbol\nu)-\boldsymbol d\bigr]
     +\mathcal R(\theta)+\mathcal P(\boldsymbol\nu).
 \label{eq:fit-functional}
\end{equation}
Here $V$ is the declared data covariance model, $\boldsymbol\nu$ denotes experimental nuisance parameters, $\mathcal P$ their constraint, and $\mathcal R$ a chosen regularizer or prior penalty. Equation~\eqref{eq:fit-functional} specifies an application workflow, not a data analysis performed in this paper. Where full covariance information is unavailable, the adopted approximation and normalization treatment must be recorded as part of the extraction.

The representation offers three linked opportunities. Correlated $x$ and $k_T$ dependence can be learned directly in the long-distance input. Perturbative operator samples and resolvent solvers can be reused as that input changes. Finally, input constraints and observable-bin sensitivities can be inspected in the same momentum-space workflow. The method therefore gives a transparent separation between what QCD fixes and what the data determine. Its quantitative benefit relative to another parameterization or numerical representation is a question for controlled closure and timing studies.

The convolution and evolution can smooth directions in parameter space, so flexibility should be matched to information content. Pseudodata generated from known admitted inputs provide a direct test of recoverable transverse structure, flavor separation, scale correlations, and regularization bias. Data spanning different $Q$, rapidity, hadron, or process channels can distinguish boundary dependence from evolution. The fitted input need not be unique for the perturbative operator and its error estimates to remain well defined. This is particularly relevant when interpreting fixed-target information outside its measured kinematic range.

\section{Observables, regulated moments, and matching}
\label{sec:observables}

\subsection{Finite-bin predictions and other TMD processes}

For the scalar Drell--Yan channel, the singular prediction at symmetric final scales is
\begin{equation}
\begin{aligned}
 W^c_{AB}(\bq;Q)
 &=\sum_qe_q^2\widehat H_{q\bar q}^{[n_H]}(Q,Q)\\
 &\quad\times
 [F^c_{q/A}\ot F^c_{\bar q/B}
       +F^c_{\bar q/A}\ot F^c_{q/B}](\bq).
\end{aligned}
 \label{eq:resummed-W}
\end{equation}
The leg arguments are $(x_A;Q,Q^2)$ and $(x_B;Q,Q^2)$. For ordinary $L^2$ legs this is bounded and has finite-bin actions. For the contact limit, the assembled symbol is admitted as described after eq.~\eqref{eq:two-leg-bound}. Fourier transformation for the proof gives the complete infrared-completed canonical CSS2 product, including unexpanded coupling and longitudinal scale dependence. Thus its bin actions equal those of that specified parent solution whenever the complete product is admitted. The boundary orders $n_H,n_B$ and the product truncation must be supplied separately. In the physical tests, Eq.~\eqref{eq:resummed-W} is assembled strictly through NLO in those boundaries, while the evolution remains unexpanded. Theorem~\ref{thm:accuracy} establishes its evolution towers; it does not supply missing finite boundary coefficients.

The same organization can be used when a parent factorization theorem supplies a TMDPDF and a TMDFF, or two TMDFFs. The physical recoil variable, fragmentation rescalings, angular projections, and hard factors then enter the relevant measurement map. The known high-order TMDFF matching functions provide the corresponding short-distance boundary inputs~\cite{EbertMistlbergerVita2021FF,LuoYangZhuZhu2021}. Gluon tensor structures have their own projectors and harmonic distributions; recent linearly polarized gluon matching calculations are pertinent inputs~\cite{ZhuLinearlyPolarized2026}. These extensions preserve the logic of the construction while requiring a channel-specific tensor and measurement analysis. Finite-mass thresholds similarly require decoupling and matching operators in addition to the fixed-flavor proof given here.

\subsection{Regulated normalization and widths}

Once the perturbative tail is included, normalization is formulated using a stated weight,
\begin{equation}
 \mathcal I_w[F_{i/h}]
     =\int\dd^2\bk\,w_K(\bk)F_{i/h}(x,\bk;\mu,\zeta).
 \label{eq:weighted-moment}
\end{equation}
For a sufficiently broad smooth family $w_K(\bk)=w(\bk/K)$, an appropriate short-distance OPE remainder bound gives
\begin{equation}
\begin{aligned}
 \mathcal I_w[F_{i/h}]
 &=\sum_j[Z^w_{i\leftarrow j}(K;\mu,\zeta)\ox f_{j/h}(\mu)](x)\\
 &\quad+\Order{(\Lambda_{\rm QCD}/K)^p}.
\end{aligned}
 \label{eq:weighted-OPE}
\end{equation}
Appendix~\ref{app:inputs} proves this sufficient case. The matching coefficient is the same transverse test action applied to $\Ccal_{ij}$. Controlled sharp-cutoff integral constructions require their endpoint analysis, as in ref.~\cite{EbertMichelStewartSun2022}.

A regulated width may be reported as
\begin{equation}
 \langle k_T^2\rangle_{w,K}
  =\frac{\int\dd^2\bk\,w_2(k_T;K)k_T^2F(x,\bk)}
         {\int\dd^2\bk\,w_0(k_T;K)F(x,\bk)},
 \label{eq:width}
\end{equation}
provided the actions exist and the denominator is nonzero. The weight, cutoff, scales, and scheme are part of its definition. A narrow soft bin probes genuine long-distance structure; it is not replaced by a collinear matching relation merely by attaching a nominal large scale to it. This is how the extraction interface distinguishes a normalized model input from a renormalized TMD observable.

\subsection{Matching the singular operator to the full spectrum}

A full-spectrum calculation combines the singular result with a nonsingular fixed-order remainder. Using a common measurement, PDFs, scales, recoil prescription, and hard normalization, define
\begin{equation}
 \dd\sigma_Y
   =\dd\sigma_{\FO}^{[m]}
       -\left[\dd\sigma_{W,\pert}^{\LP}\right]_{a^m}.
 \label{eq:Y-subtraction}
\end{equation}
The subtraction is the perturbative leading-power expansion. In particular, it is not an expansion of a fitted long-distance smearing. Writing
$\Delta\dd\sigma_{W,\NP}=\dd\sigma_{W,\rm full}-\dd\sigma_{W,\pert}$ gives
\begin{equation}
\begin{aligned}
 \dd\sigma_{\rm matched}
 &=\dd\sigma_{\FO}^{[m]}
       +\dd\sigma_{W,\pert}^{\rm resum}
                     -[\dd\sigma_{W,\pert}^{\LP}]_{a^m}\\
 &\quad+\Delta\dd\sigma_{W,\NP}
       +\Delta\dd\sigma_{\rm rem}.
\end{aligned}
 \label{eq:matched-spectrum}
\end{equation}
Here $\Delta\dd\sigma_{\rm rem}$ denotes only power effects not already present in the fixed-order remainder or an explicit recoil improvement. The universal TMD large-$k_T$ coefficient $\Ccal\ox f$ and the process-specific nonsingular remainder are different ingredients.

Consistent resummation turnoff is applied to the retained prediction and its subtraction as part of the process definition. Inclusive rates and the region $q_T\sim Q$ are tested using this matched result. Precision fiducial measurements can have recoil-induced power corrections whose first power differs from an inclusive azimuthally averaged spectrum~\cite{EbertMichelStewartTackmann2021}. Accordingly, the leading-power logarithmic label and the treatment of measurement-induced powers are specified separately.

Fixed-order and matched implementations such as DYTurbo, SCETlib, and CuTe--MCFM provide inputs and independent benchmarks for this stage~\cite{CamardaEtAlDYTurbo2020,CamardaCieriFerrera2021,BillisMichelTackmann2025,NeumannCampbell2023}. Their use does not modify the definition of an individual universal TMD. It tests the physical assembly, subtraction, and measurement rather than merely the algebra of its transverse basis.

\subsection{Uncertainty organization}

The formulation separates four types of numerical and theoretical control. Renormalization and rapidity-scale variations probe omitted perturbative terms, with final rapidity products constrained by the process. Matching and turnoff variations probe the transition to fixed order. Variations of the infrared separation are accompanied by the corresponding nonperturbative-boundary and CS-kernel treatment. Finally, RG quadrature, longitudinal interpolation, polynomial approximation, elliptic-grid, and bin-integration errors form a numerical budget.

These are categories, not an instruction to add independent errors in quadrature. Their correlations or envelope prescription are part of a numerical analysis. Experimental, PDF, and long-distance-model uncertainties enter the later extraction and remain distinct from the perturbative accuracy assignment. A comparison between momentum-space and impact-parameter-space fits is most informative when the physical ingredients and infrared convention are held fixed, so differences associated with parameterization and numerical representation can be resolved.

\section{Physical implementation and computational validation}
\label{sec:verification}
\label{sec:validation}

\subsection{Sourced coefficients and strict-NLO assembly}
\label{sec:physical}

The physical realization uses the quark CSS2 evolution vector with four-loop
beta/cusp and three-loop noncusp/CS-boundary inputs. The repository records the
primary source, equation location, convention conversion and hashes for each
coefficient. The four-loop cusp is compared with an independently written
expression \cite{HKM2020,MPS2020} at $n_f=3,4,5,6$; the largest absolute
disagreement is below $2.4\times10^{-11}$. Common source coefficients are physical
inputs to both numerical paths, not independently fitted quantities.

Appendix~\ref{app:one-loop} gives the one-loop matching coefficients, plus
convention, hard normalization and complete fixed-order Drell--Yan check.
At a canonical boundary, write a leg's longitudinal factor as
$B_A=B_A^{(0)}+a_c C_{qq,A}^{(1)}+a_c C_{qg,A}^{(1)}$,
and similarly for $B$. The two-leg boundary is assembled as
\begin{align}
 {\cal P}^{[1]}_{AB}={}&B_A^{(0)}B_B^{(0)}
       +a_Qh_q^{(1)}B_A^{(0)}B_B^{(0)}\notag\\
 &+a_c\bigl(C_{qq,A}^{(1)}+C_{qg,A}^{(1)}\bigr)B_B^{(0)}
       \notag\\
 &+a_c B_A^{(0)}\bigl(C_{qq,B}^{(1)}+C_{qg,B}^{(1)}\bigr).
 \label{eq:strict-components}
\end{align}
The notation shows one flavor orientation before summing charge squares and its
charge-conjugate partner; the longitudinal actions include the appropriate PDFs.
The canonical coupling $a_c=a(\mu_c(s))$ and final coupling $a_Q=a(Q)$ are both
unexpanded. Multiplication by the two evolution factors leaves six linear
components: Born, hard, $qq$ on either leg, and $qg$ on either leg. Hard--OPE and
OPE--OPE products are omitted consistently with strict-NLO boundary counting.

For general final $\mu$ the one-loop hard insertion is
\begin{equation}
\begin{aligned}
 \delta H^{(1)}&=a(\mu)C_F
 \left[-16+\frac{7\pi^2}{3}+6L_Q-2L_Q^2\right],\\
 L_Q&=\ln\frac{Q^2}{\mu^2}.
\end{aligned}
 \label{eq:hard-general}
\end{equation}
Its scale derivative cancels the two TMD virtuality anomalous dimensions at
this order when $\zeta_A\zeta_B=Q^4$. The splitting-function derivative in the
matching cancels PDF evolution at the common order. These checks test the signs,
longitudinal normalization and assembly in addition to the transverse operator.

The public verification registry contains 14 required suites.  They cover
fixed-order distribution algebra, longitudinal convolutions, RG composition,
running-coupling expansions, contact-preserving weak actions, ordered matrix
fixtures, resolvent recurrences, formal tower checks, and the sourced physical
canonical evolution input.  The recorded audit passes all 14 suites; the software
contract layer passes 121 tests.  Test counts are not added into a theorem
certificate.

Here ``independent reference'' means a separately coded numerical path, with
different quadrature and evolution implementations.  Both paths were developed
within the same computational effort.  Their agreement does not constitute
external scientific review or an independent determination of the shared
perturbative coefficients.

\subsection{Non-Gaussian continuum comparison}

We use the normalized Student-type radial input
\begin{equation}
 f(k;\lambda,p)=\frac{p-1}{\pi\lambda^2}
 \left(1+\frac{k^2}{\lambda^2}\right)^{-p},
 \quad p>1.
\end{equation}
The ordinary-input benchmark uses $\lambda=1.2~\mathrm{GeV}$ and $p=3$. This density has an algebraic tail and an analytic Hankel transform proportional to
$(\lambda b)^{p-1}K_{p-1}(\lambda b)$.  Differentiating this expression gives an
independent analytic $\partial f/\partial\lambda$ reference.

The continuum calculation keeps $M_\infty f$ analytically and integrates only
the compact nonconstant remainder by adaptive Bessel quadrature.  The native
calculation uses the radial Laplacian and Chebyshev resolvent action.  Seven
finite $q_T$ bins cover 0--8~GeV.  The final calculation has degree 1024,
$\Delta k=0.0125$~GeV, and radius 1280~GeV.  The radius is numerical and does
not assert physical support to that momentum.

\begin{figure*}[t]
 \centering
 \includegraphics[width=0.97\linewidth]{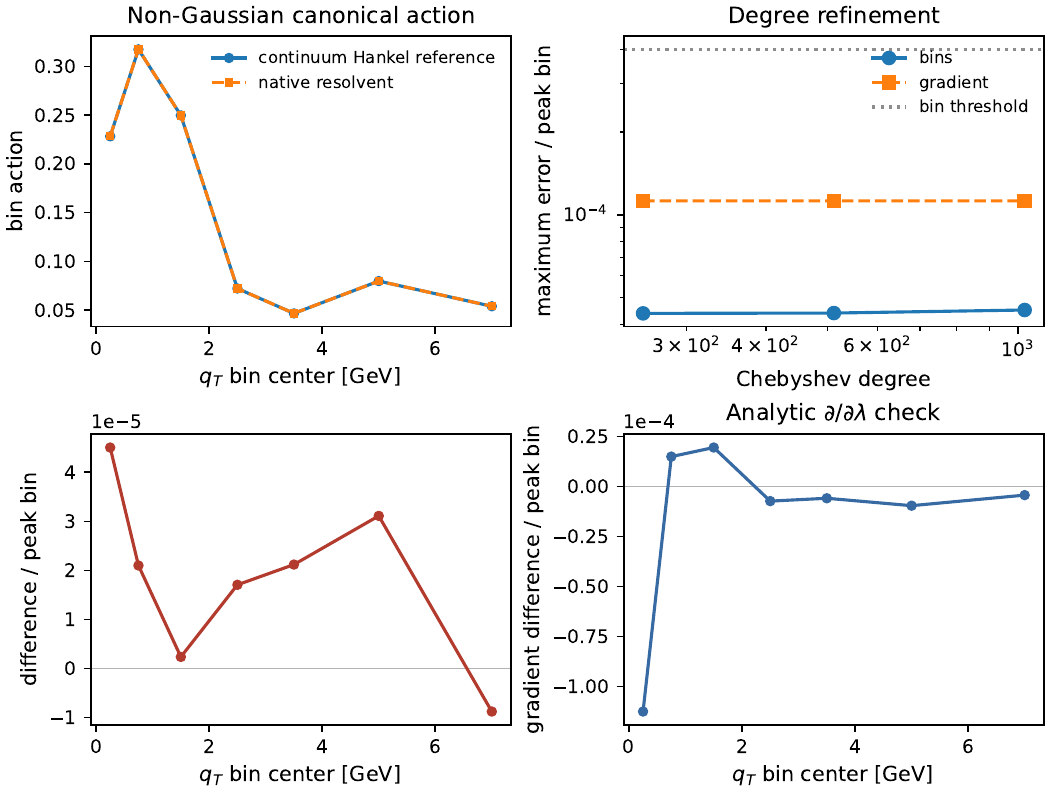}
 \caption{Independent continuum validation for the physical canonical N$^3$LL
 evolution symbol acting on a non-Gaussian algebraic-tail input.  The lower
 panels show differences normalized by the peak reference bin.  The gradient is
 with respect to the input scale $\lambda$. Endpoint-preserving polynomial-degree refinement is shown separately from continuum discretization.}
 \label{fig:continuum}
\end{figure*}

Figure~\ref{fig:continuum} shows the ordinary-bin and derivative comparison. The maximum native/reference bin discrepancy is $4.51\times10^{-5}$ of the peak
bin and the maximum analytic-gradient discrepancy is $1.12\times10^{-4}$.
The last degree, spacing, and box changes are $2.90\times10^{-5}$,
$1.32\times10^{-4}$, and $2.17\times10^{-5}$, respectively.  Adding these to
the final discrepancy gives a conservative observed bin budget
$2.28\times10^{-4}$, below the registered $4\times10^{-4}$ criterion.
The adaptive reference error is below $3\times10^{-13}$.

\subsection{Same-prescription physical two-leg validation}
\label{sec:canonicalphysical}

To connect the evolution theorem directly to a physical boundary, we combine
the uncapped canonical operator with NNPDF4.0 NNLO member~0~\cite{NNPDF2022},
accessed through LHAPDF6~\cite{LHAPDF2015}, five massless quark
flavors, charge sums, both beam orientations, and all six strict-NLO components.
The three endpoints are $Q=8$, $20$, and $91.1876$~GeV, with
$(x_A,x_B)=(0.08,0.20)$, $(0.15,0.30)$, and $(0.02,0.07)$, respectively.
These are photon-channel benchmarks; the highest scale does not imply that a
$Z$ contribution has been included.  The coupling is independently evolved at
four loops with $n_f=5$ from $\alpha_s(91.1876~\mathrm{GeV})=0.1180024$.

The matching coordinate is $b_{\rm eval}=b_0/\mu_c(s)$ and is therefore the
actual $b$ in the canonical perturbative region.  The completed coordinate
defines the infrared boundary elsewhere.  No $Q$-dependent ultraviolet cap is
introduced.  The low- and high-scale cases use two Student legs with powers
$p=3$ and $q=4$ and scales
$\lambda_A=0.9+0.8x_A$~GeV and $\lambda_B=1.2+0.6x_B$~GeV; the middle case
uses a Gaussian pair.  Native convolution of the Student legs is evaluated
with the momentum-space Feynman-parameter identity derived in Appendix~\ref{app:physical-reference}.

The reference shares the physical coefficient numbers and official PDF provider,
but imports none of the production RG, matching, flavor-sum, or physical-symbol
routines.  It independently integrates the full endpoint anomalous dimension
with RK45, evaluates matching in $\ln\xi$ instead of $z$, and uses adaptive
Bessel quadrature with the product of two analytic input transforms.  Adaptive
longitudinal integrals check the reference cache.  Separate tolerances are
refined in the RG calculation.  A logarithmic derivative with respect to
$\lambda_A$ and an explicit contact input accompany the ordinary-bin checks.

\subsection{Finite PDF tables and the uncapped ultraviolet limit}

The source PDF table ends at $\mu_{\max}=10^5$~GeV.  The production nodes remain
inside that table.  We bound the unresolved interval $b<b_0/\mu_{\max}$ for the
explicit sufficient class
\begin{equation}
 |f_j(x,\mu)|\le |f_j(x,\mu_{\max})|
 [1+\ln(\mu/\mu_{\max})]^4,
 \qquad \mu>\mu_{\max},
 \label{eq:pdfuvclass}
\end{equation}
including the arguments appearing inside longitudinal convolutions.  This is an
admitted input condition, not knowledge supplied by the finite table.  Absolute
matching integrals and monotone ultraviolet RG decay give a component-wise bin
bound.  The largest such bound is $1.67\times10^{-6}$ of the peak bin; it remains
in the error budget.

We also complete the reference input explicitly with two sensitivity fixtures:
PDFs constant above the table, or multiplied there by
$[1+\ln(\mu/\mu_{\max})]^4$.  The coupling and canonical RG are uncapped in both
cases.  These fixtures are not asserted to solve DGLAP above the table.  Resolving
their ultraviolet integrals leaves an integration/tail error below
$10^{-17}$ of the peak; changing between them shifts any bin by at most
$1.54\times10^{-6}$.  This input dependence is distinct from numerical error.
The achieved broader class bound is $1.67\times10^{-6}$, passing the declared
$2\times10^{-6}$ criterion.  This is a numerical tolerance, with no identified
physical significance attached to that value.

\begin{figure*}[t]
 \centering
 \includegraphics[width=\linewidth]{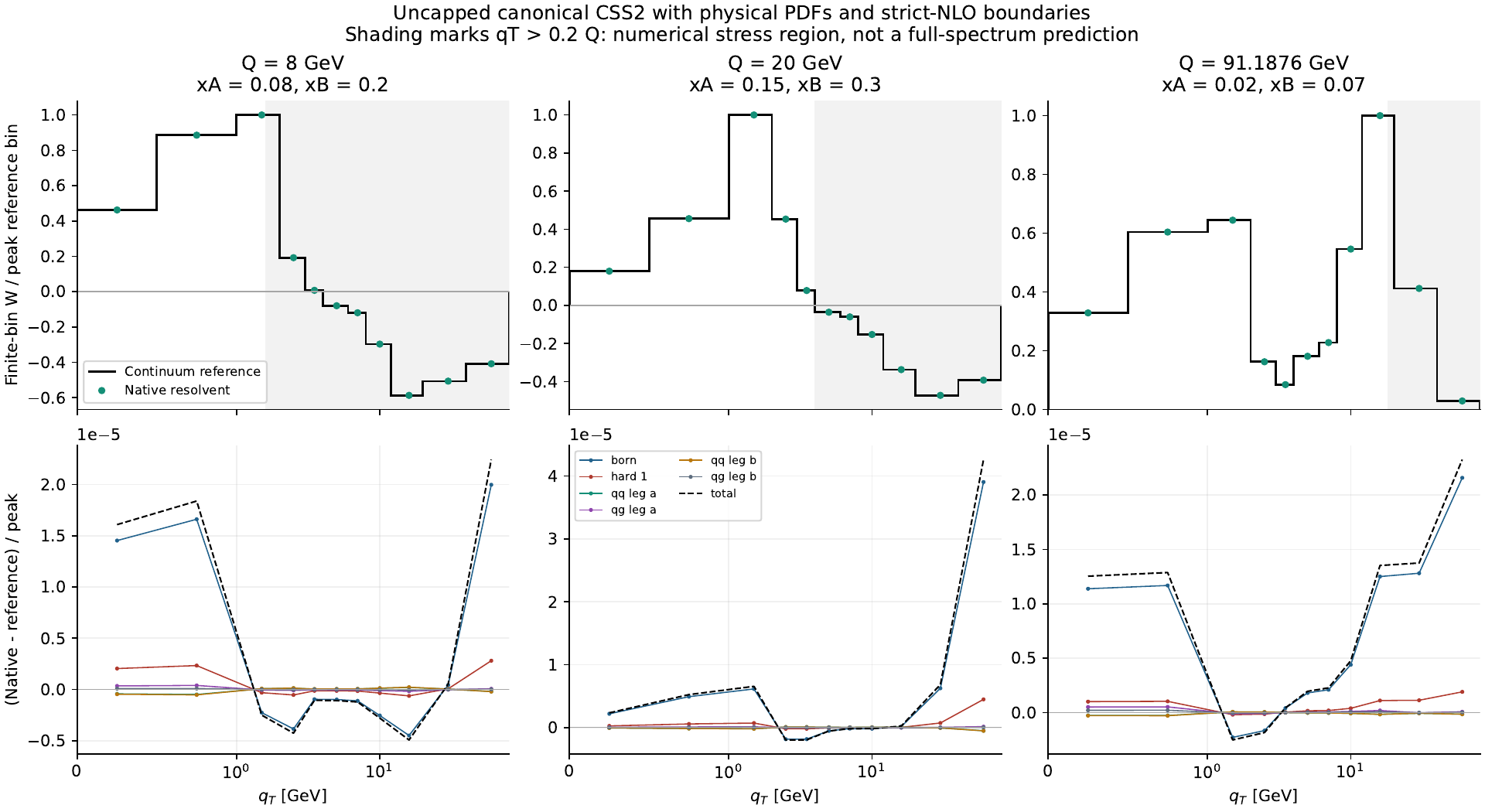}
 \caption{Physical PDFs and all strict-NLO components evaluated with the same
 uncapped canonical prescription in both calculations.  The upper panels show
 finite-bin values; the lower panels retain each signed component error.
 Shading denotes $q_T>0.2Q$, used as a numerical stress region rather than a
 complete physical spectrum.}
 \label{fig:canonicalphysical}
\end{figure*}

\begin{table*}[t]
\caption{Errors normalized to the peak reference total in each case, including
 the broader PDF continuation bound of Eq.~\eqref{eq:pdfuvclass}.  The gradient
 is with respect to $\ln\lambda_A$.  The observed budget adds the bin discrepancy,
 degree, spacing and box changes, continuum quadrature error, and ultraviolet
 class bound; it is not a certified global supremum.}
 \label{tab:physicalaudit}
 \centering
 \setlength{\tabcolsep}{9pt}
 \begin{tabular}{rrrrr}
 \toprule
 $Q$ [GeV] & Bin error & Gradient error & Contact error & Observed budget\\
 \midrule
 8       & $2.25\times10^{-5}$ & $2.56\times10^{-5}$ & $2.39\times10^{-5}$ & $2.63\times10^{-4}$\\
 20      & $4.27\times10^{-5}$ & $6.12\times10^{-6}$ & $4.08\times10^{-5}$ & $3.47\times10^{-4}$\\
 91.1876 & $2.50\times10^{-5}$ & $2.64\times10^{-5}$ & $2.14\times10^{-5}$ & $1.85\times10^{-4}$\\
 \bottomrule
 \end{tabular}
\end{table*}

\subsection{Refinement budget and acceptance}

\begin{figure*}[t]
 \centering
 \includegraphics[width=0.9\textwidth]{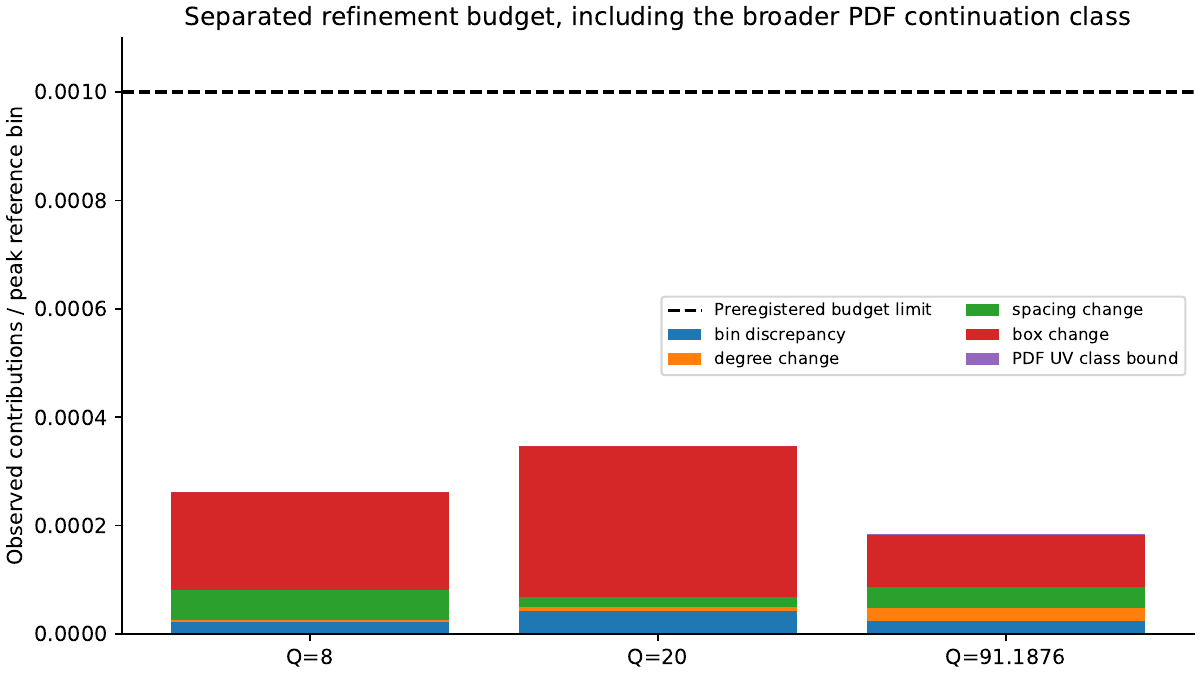}
 \caption{Separated numerical contributions to the physical finite-bin budget.
 The changes in polynomial degree, radial spacing and outer radius are measured
 independently. The conditional UV input bound is included in the primary
 errors. Their observed sum controls this benchmark; it is not an experimental
 uncertainty or a certified supremum over all admissible inputs.}
 \label{fig:budget}
\end{figure*}

Figure~\ref{fig:canonicalphysical} shows the physical comparison, and Table~\ref{tab:physicalaudit} summarizes its errors. The degree, spacing, and box are refined separately; Fig.~\ref{fig:budget} displays their contributions.  The accepted high-scale
calculation uses radius 5120~GeV, degree 4096, and $\Delta k=0.0125$~GeV,
with lower-degree, coarser-spacing, and smaller-box runs supplying the refinement
comparisons.  The other cases use radius 1280~GeV and degree 2048.  These are numerical domains,
not physical momentum support.  The fixed acceptance limits are
$5\times10^{-4}$ for bins, components and contacts, $10^{-3}$ for gradients,
and $10^{-3}$ for the observed budget.  All pass, including the broader
continuation-class errors.  The data package records the accepted results,
their refinements, and the separate completed-input comparisons.

\section{Bounded proton--proton Drell--Yan and inverse demonstrations}
\label{sec:demo}

The process-level demonstrator uses 41 proton--proton E866 bins transcribed
from Ref.~\cite{Webb2003} at
$\sqrt{s}=38.8$~GeV, selected with $Q_{\rm low}\ge6.2$~GeV and
$q_{T,\rm high}/Q_{\rm low}\le0.2$.  The source manifest pins the thesis source,
table lines, units, bounds, values, and statistical and systematic errors.  The
calculation uses NNPDF4.0 NNLO member~0, independent four-loop fixed-$n_f=5$
running, photon charges, the six strict-NLO components, and integration over the
reported mass, $x_F$, and transverse bins.  The profile keeps perturbative scales
above 4.99~GeV; this avoids threshold crossing but is not a massive-quark
calculation.

This demonstrator uses a separately specified capped process profile.  Its
agreement tests the physical workflow under that prescription.  It supplies no
additional proof of the uncapped canonical theorem; the direct connection to
physical matching is provided by Sec.~\ref{sec:canonicalphysical}.

The native forward benchmark agrees with independent PDF convolutions, RG
integration, and Bessel quadrature.  At the final refinement its worst bin error
is $2.26\times10^{-6}$, its component error is $2.12\times10^{-6}$ of the row
total, and its scaled gradient error is $3.78\times10^{-5}$.  A two-parameter
Gaussian pseudodata closure recovers injected values with biases
$1.21\times10^{-6}$ and $-9.7\times10^{-8}$; the maximum held-out relative
prediction error is $8.2\times10^{-6}$.

\begin{figure*}[t]
 \centering
 \includegraphics[width=0.97\linewidth]{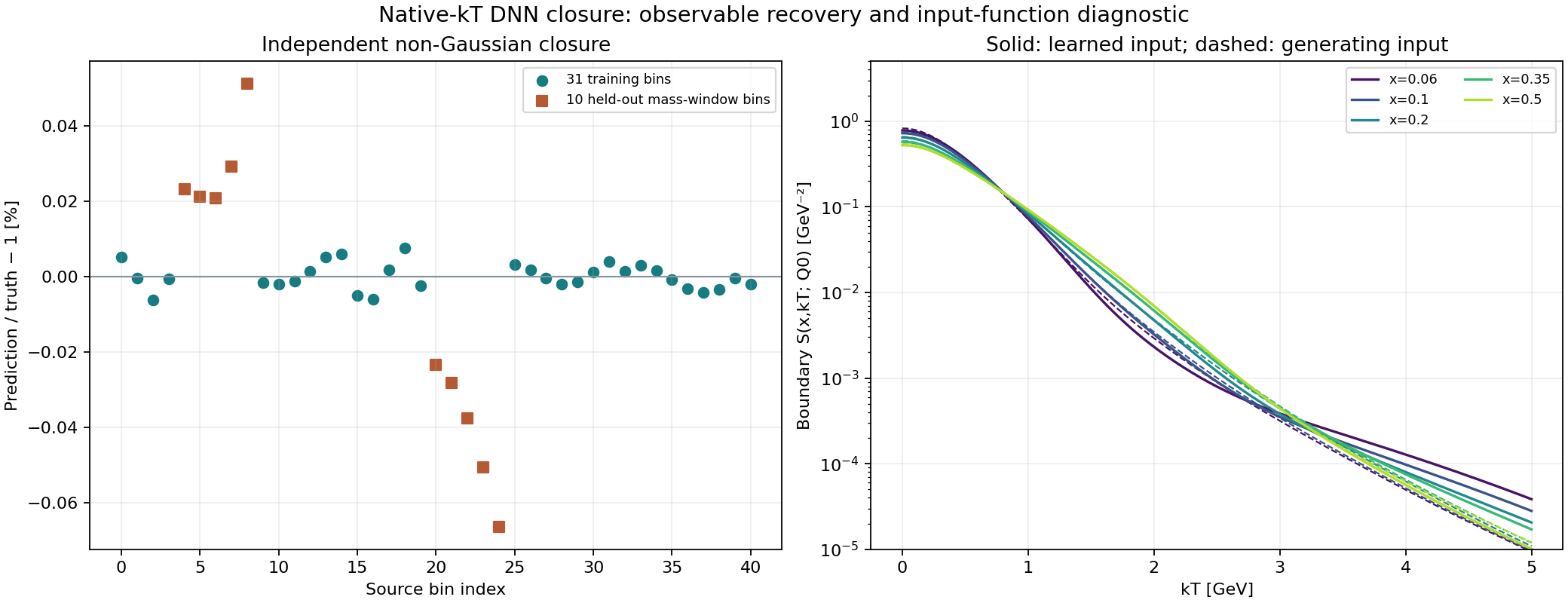}
 \caption{Restricted native-$\kt$ neural closure.  The left panel shows relative
 observable errors, including a held-out mass window; the right panel compares
 learned and generating boundary profiles.  Agreement of the observables within
 this six-width family does not establish unique or unrestricted recovery of the
 input function.}
 \label{fig:closure}
\end{figure*}

A restricted native neural pilot replaces the single Gaussian with six fixed
radial widths and learns $x$-dependent softmax weights using a
$1$--$12$--$12$--$12$--$6$ network.  Production training uses precomputed native
operator responses and analytic gradients.  In a non-Gaussian, $x$-correlated
closure, all ten bins in one mass window are held out.  The accepted seed reaches
a maximum training error $7.6\times10^{-5}$ and held-out error
$6.7\times10^{-4}$, while the native/refined discrepancy remains
$6.9\times10^{-6}$.  Figure~\ref{fig:closure} shows the closure in observable and input space. This demonstrates differentiable inverse use of the
operator.  Because the generator belongs to the same six-width radial family,
it does not establish unrestricted functional recovery.

The actual-data result is deliberately secondary.  The original E866 mean-energy
column is inconsistent with the stated bin kinematics for 253 of 279 transcribed
rows, and the exact finite-bin conversion to the reported invariant is not fully
documented.  A conditional neural fit using the preserved tabulated-mean
estimator gives point plus normalization-prior $\chi^2=21.68$ for 41 bins and
$N=1.024$, but no parameter-counting $\chi^2/{\rm dof}$ or extraction claim is
assigned.  The fixed nonperturbative CS coefficient is inherited from a
conditional Gaussian diagnostic.  These restrictions make the result a workflow
demonstration rather than TMD phenomenology.

\section{Scope, applications, and conclusions}
\label{sec:conclusions}

We have presented a self-contained native-$k_T$ realization of the Collins/CSS2
singular TMD operator, from its fixed-order distribution algebra to an
unexpanded evolution and a constructive numerical action. The distinction
between a parent canonical boundary and a momentum-disk boundary fixes the
finite contacts; ordered longitudinal recurrences retain flavor structure.
An endpoint-independent infrared completion and asymptotically free running
give admitted two-scale evolution on a common logarithmically weighted domain.
Complete canonical legs are bounded, including matched longitudinal inputs
under the stated growth conditions.

The resolvent representation turns that construction into momentum-space
convolutions or elliptic solves. Its contact extension and finite-bin bounds
make precise which distributions and observables are admitted. The all-order
degree argument establishes the four exponent towers of $N^3\mathrm{LL}$
evolution. We have also specified higher-order coefficient projection and
product assembly, so the formalism can be extended when the necessary physical
boundary coefficients have been ingested and checked.

The physical implementation reported here has strict-NLO hard/OPE boundaries.
Its uncapped three-scale benchmark uses actual PDFs, both legs, six linear
components, algebraic-tail inputs, contacts and analytic derivatives. With the
declared PDF ultraviolet continuation class included, native and continuum bin
actions agree below $4.27\times10^{-5}$ of the peak. Separate refinements and
explicit UV completions determine the numerical acceptance. The published
source, accepted records and commands make those claims reproducible.

The long-distance input interface permits joint $x$ and $k_T$ dependence in a
network, spline or analytic function, with evolution supplied by the operator.
Localized smearing, a separate universal nonperturbative CS term, and regulated
moments have different physical roles. The bounded forward map controls
derivatives with respect to input parameters. A conditional E866 calculation
and restricted neural closure demonstrate that interface without establishing
general functional recovery or a precision global extraction.

The result imports the parent factorization theorem and is proved for the stated
scalar, fixed-flavor, infrared-completed domain. A numerical $\Nthree$ spectrum,
massive thresholds, tensor/fragmentation channels and a complete matched $W+Y$
prediction require the additional ingredients identified above. Their
prescriptions can be described within this framework without implying that
those calculations have been completed. The matched measurement, rather than
the uncapped singular term alone, controls inclusive normalization and the
large-$q_T$ spectrum.

A separate phenomenological project will compare a joint-$(x,k_T)$ FiLM
extraction with the $b_T$-space analysis of Ref.~\cite{FernandoKeller2026}, using
identical data and matched statistical treatment. That comparison will test
representation-dependent inference after the present method is fixed. The
first paper's result is the complete prescription, its accuracy domain and
its computational validation.

\section*{Data and code availability}

The scientific software and validation package are available at the immutable
revision \href{https://github.com/uva-spin/k-space/tree/04cf2ff882067c047216fb09e1f4df04ff78acc0}{\texttt{04cf2ff}}
of Ref.~\cite{KSpaceSoftware}.
The package contains the numerical reports behind the
figures and tables, the coefficient and PDF source manifests, refinement data,
and checksums.  Its dated evidence snapshot is
\nolinkurl{validation/reference/2026-10-01}; the reproduction guide is
\nolinkurl{docs/REPRODUCE_METHOD_PAPER.md}.  External PDF tables and primary-source
archives are obtained using the supplied acquisition scripts and verified
against pinned hashes.

\appendix
\section{Explicit one-loop coefficients and the Drell--Yan check}
\label{app:one-loop}

The following formulas fix the normalization used throughout. The collinear splitting functions obey
\begin{equation}
 \frac{\dd f}{\dd\ln\mu^2}
     =\sum_{r\ge1}\left(\frac{\alpha_s}{2\pi}\right)^r
                      P^{(r-1)}\ox f.
 \label{eq:DGLAP-convention}
\end{equation}
For QCD, $T_F=1/2$, $C_A=N_c$, and $C_F=(N_c^2-1)/(2N_c)$. Define
\begin{align}
 P_{qq}^{(0)}(z)&=C_F\left[(1+z^2)\left(\frac{1}{1-z}\right)_+
                         +\frac32\delta(1-z)\right],\notag\\
 P_{qg}^{(0)}(z)&=T_F[z^2+(1-z)^2].
 \label{eq:splittings-one}
\end{align}
The plus prescription in the first term acts on $1/(1-z)$; this specifies the numerator and contact convention unambiguously.

The CSS2 canonical quark matching coefficients in $C=C^{(0)}+aC^{(1)}+\cdots$ are~\cite{CollinsRogers2017}
\begin{align}
 c_{q\leftarrow q}^{(1)}(z)
      &=C_F\left[2(1-z)-\frac{\pi^2}{6}\delta(1-z)\right],\notag\\
 c_{q\leftarrow g}^{(1)}(z)&=4T_Fz(1-z).
 \label{eq:canonical-one}
\end{align}
The other quark-flavor and antiquark-to-quark entries vanish at this order. The antiquark row follows by charge conjugation. Restoring scale logarithms gives
\begin{equation}
\begin{aligned}
 \widetilde C_{q\leftarrow q}^{(1)}
 &=c_{q\leftarrow q}^{(1)}-2L_bP_{qq}^{(0)}\\
 &\quad+C_F\delta(1-z)[(3-2\ell_\zeta)L_b-L_b^2],\\
 \widetilde C_{q\leftarrow g}^{(1)}
 &=c_{q\leftarrow g}^{(1)}-2L_bP_{qg}^{(0)} .
\end{aligned}
 \label{eq:one-b-coefficients}
\end{equation}
The momentum-space coefficients are consequently
\begin{equation}
\begin{aligned}
 \Ccal_{q\leftarrow q}^{(1)}(z,\bk)
 &=c_{q\leftarrow q}^{(1)}(z)\deltaT(\bk)
     +2P_{qq}^{(0)}(z)\Ln_0(\bk,\mu)\\
 &\quad-C_F(3-2\ell_\zeta)\delta(1-z)\Ln_0(\bk,\mu)\\
 &\quad+2C_F\delta(1-z)\Ln_1(\bk,\mu),\\
 \Ccal_{q\leftarrow g}^{(1)}(z,\bk)
 &=4T_Fz(1-z)\deltaT(\bk)+2P_{qg}^{(0)}(z)\Ln_0(\bk,\mu).
\end{aligned}
 \label{eq:one-k-coefficients}
\end{equation}
The explicit $T_F$ simply restores the QCD trace factor in the finite $2z(1-z)$ coefficient of the parent convention.

For the CS kernel,
\begin{equation}
\begin{aligned}
 D_i^{(1)}&=\Gamma_0^iL_b,\qquad
 \Dcal_i^{(1)}=\Gamma_0^i\Jdist_0=-\Gamma_0^i\Ln_0,\\
 \Gamma_0^q&=4C_F,\qquad \Gamma_0^g=4C_A .
\end{aligned}
 \label{eq:one-D}
\end{equation}
The one-loop noncusp coefficient for the quark TMD is $\gamma_0^q=6C_F$. At fixed $\zeta$, eq.~\eqref{eq:one-b-coefficients} therefore satisfies
\begin{align}
 \partial_{\ln\mu}\widetilde C_{q\leftarrow j}^{(1)}
  &=(6C_F-4C_F\ell_\zeta)\delta_{qj}\delta(1-z)-4P_{qj}^{(0)},\notag\\
 \partial_{\ln\sqrt\zeta}\widetilde C_{q\leftarrow j}^{(1)}
  &=-4C_FL_b\,\delta_{qj}\delta(1-z).
 \label{eq:one-RG-check}
\end{align}
These equations fix both the rapidity sign and the factor relating the standard splitting kernel to evolution in $\ln\mu$.

At $\mu=Q$ the reduced Drell--Yan hard coefficient is~\cite{CollinsRogers2017}
\begin{equation}
\begin{aligned}
 \widehat H_{q\bar q}(Q,Q)&=1+ah_q^{(1)}+\Order{a^2},\\
 h_q^{(1)}&=C_F\left(-16+\frac{7\pi^2}{3}\right).
\end{aligned}
 \label{eq:one-hard}
\end{equation}
To give the full flavor structure compactly, define
\begin{equation}
\begin{aligned}
 b_{qq}&=2P_{qq}^{(0)}-3C_F\delta(1-z),\\
 b_{qg}&=2P_{qg}^{(0)},\\
 (bf)_{q/h}&=b_{qq}\ox f_{q/h}+b_{qg}\ox f_{g/h},\\
 (cf)_{q/h}&=c_{q\leftarrow q}^{(1)}\ox f_{q/h}
                   +c_{q\leftarrow g}^{(1)}\ox f_{g/h},\\
 \Phi_q^{(0)}&=f_{q/A}f_{\bar q/B}+f_{\bar q/A}f_{q/B}.
\end{aligned}
 \label{eq:luminosities-one}
\end{equation}
All PDFs are evaluated at fixed scale $Q$ and the appropriate $x_A,x_B$. Let
\begin{equation}
\begin{aligned}
 \Phi_q^{b,A}&=(bf)_{q/A}f_{\bar q/B}+(bf)_{\bar q/A}f_{q/B},\\
 \Phi_q^{b,B}&=f_{q/A}(bf)_{\bar q/B}+f_{\bar q/A}(bf)_{q/B}.
\end{aligned}
 \label{eq:b-insertions}
\end{equation}
and define $\Phi_q^{c,A},\Phi_q^{c,B}$ by replacing $b$ with $c$. At symmetric final rapidities the one-loop singular result is
\begin{equation}
\begin{aligned}
 W^{(0+1)}_{AB}(\bq)
 &=\sum_qe_q^2\Bigl\{
       \Phi_q^{(0)}\deltaT(\bq)
       +4aC_F\Phi_q^{(0)}\Ln_1(\bq,Q)\\
 &\quad+a(\Phi_q^{b,A}+\Phi_q^{b,B})\Ln_0(\bq,Q)\\
 &\quad+a(h_q^{(1)}\Phi_q^{(0)}+\Phi_q^{c,A}+\Phi_q^{c,B})
                \deltaT(\bq)\Bigr\}\\
 &\quad+\Order{a^2}.
\end{aligned}
 \label{eq:one-W-full}
\end{equation}
Integrating this expression with eq.~\eqref{eq:plus-bins} tests the complete first bin and the ordinary annular bins. Isolating the Sudakov diagonal contribution from the separately displayed longitudinal splitting insertions gives
\begin{equation}
\begin{aligned}
 &\pi a[4C_F\Ln_1(\bq,Q)-6C_F\Ln_0(\bq,Q)]\\
 &\quad=\frac{\alpha_s}{\pi q_T^2}
        \left[C_F\ln\frac{Q^2}{q_T^2}-\frac32C_F\right],
        \qquad q_T>0.
\end{aligned}
 \label{eq:AB-check}
\end{equation}
The factor $\pi$ converts the radial $\dd^2\bq$ distribution to the $\dd q_T^2$ spectrum. It reproduces $A_q^{(1)}=C_F$ and $B_q^{(1)}=-3C_F/2$ with the declared splitting convention and fixed-PDF coefficient comparison.

\section{Ordered matching reconstruction and finite boundary translations}
\label{app:matching}

\subsection{Collins--Soper logarithms through three loops}

The kernel RG equation determines the logarithms once the canonical constants are supplied. With $L=L_b$ and $a=a(\mu)$,
\begin{equation}
\begin{aligned}
 D_i^{[3]}(L;\mu)={}&a\Gamma_0^iL\\
 &+a^2\left[\frac{\beta_0\Gamma_0^i}{2}L^2+\Gamma_1^iL+d_i^{(2)}\right]\\
 &+a^3\biggl[\frac{\beta_0^2\Gamma_0^i}{3}L^3\\
 &\qquad+\left(\beta_0\Gamma_1^i+\frac{\beta_1\Gamma_0^i}{2}\right)L^2\\
 &\qquad+(\Gamma_2^i+2\beta_0d_i^{(2)})L+d_i^{(3)}\biggr].
\end{aligned}
 \label{eq:D-three-polynomial}
\end{equation}
At two loops,
\begin{align}
 \Gamma_1^i&=4C_i\left[\left(\frac{67}{9}-\frac{\pi^2}{3}\right)C_A
                       -\frac{20}{9}T_Fn_f\right],\notag\\
 d_i^{(2)}&=C_i\left[C_A\left(\frac{808}{27}-28\zeta_3\right)
                       -\frac{224}{27}T_Fn_f\right],
 \label{eq:two-loop-inputs}
\end{align}
with $C_i=C_F$ for quarks and $C_i=C_A$ for gluons at this order~\cite{CollinsRogers2017,LiZhu2017}. The four-loop cusp uses its full color structure. Differentiating eq.~\eqref{eq:D-three-polynomial} with the running coupling gives $2\Gamma_i$ through the common third order. The unexpanded kernel in eq.~\eqref{eq:D-resummed} instead satisfies eq.~\eqref{eq:D-exact-RG} exactly within the retained four-loop system.

The polynomial is used for fixed-order matching and checks. Literal global exponentiation of its cubic term would grow as $\exp(c|\ln b_T|^3)$ at one spectral endpoint for a nonzero real rapidity interval. The RG-improved construction avoids that extrapolation: it integrates the coupling before exponentiation and has the logarithmic-power behavior in eq.~\eqref{eq:D-asymptotic}. This is why fixed-order projection and unexpanded evaluation are distinct steps.

\subsection{Ordered flavor recurrence}

Write the longitudinal matrices as
\begin{equation}
\begin{aligned}
 C&=\sum_{n\ge0}a^n C_n(L,\ell),\qquad C_0=\mathbf1,\\
 D&=\sum_{r\ge1}a^rD_r(L),\\
 P_\mu&=\sum_{r\ge1}a^r p_r,\qquad p_r=2^{r+1}P^{(r-1)} .
\end{aligned}
 \label{eq:matrix-conventions}
\end{equation}
The unit includes $\delta_{ij}\delta(1-z)$. The anomalous dimensions and $D_r$ are diagonal in the outgoing TMD row, whereas $p_r$ acts on the right. Products of longitudinal matrices below include $\ox$.

\begin{proposition}[Logarithmic reconstruction with ordered flavor products]
\label{prop:recurrence}
Let $\bar C_n(L)=C_n(L,0)$ and supply $\bar C_n(0)=c_n(z)$ and $D_n(0)=d_n$. Then
\begin{align}
D_n'(L)&=\Gamma_{n-1}
       +\sum_{r=1}^{n-1}(n-r)\beta_{r-1}D_{n-r}(L),
 \label{eq:D-recurrence}\\
 2\bar C_n'(L)&=\sum_{r=1}^{n}(\gamma_{r-1}-D_r)\bar C_{n-r}\notag\\
 &\quad-\sum_{r=1}^{n}\bar C_{n-r}\ox p_r\notag\\
 &\quad+2\sum_{r=1}^{n-1}(n-r)\beta_{r-1}\bar C_{n-r},
 \label{eq:C-recurrence}\\
 C(a,L,\ell)&=e^{-\ell D(a,L)/2}\bar C(a,L).
 \label{eq:C-rapidity}
\end{align}
The equations determine all logarithms uniquely order by order from the finite canonical boundaries.
\end{proposition}
\begin{proof}
At fixed $b_T$ and $\zeta$, the scale derivative is
$2\partial_L-2\partial_\ell+\beta(a)\partial_a$.
The coefficient equations are $2\partial_\ell C=-DC$ and
$\dd C/\dd\ln\mu=\gamma_F C-C\ox P_\mu$. Substituting the first into the second and setting $\ell=0$ gives eq.~\eqref{eq:C-recurrence} after expansion in $a$. The running term has a positive sign when moved to the right-hand side. Applying the same comparison to $\dd D/\dd\ln\mu=2\Gamma$ gives eq.~\eqref{eq:D-recurrence}. Integration from $L=0$ determines each coefficient successively. Equation~\eqref{eq:C-rapidity} is the diagonal solution of the rapidity equation, interpreted here as a formal series to the specified order.
\end{proof}

For example,
\begin{align}
 2\bar C_3'={}&(\gamma_2-D_3)\mathbf1-p_3
       +(\gamma_1-D_2)\bar C_1-\bar C_1\ox p_2\notag\\
       &+(\gamma_0-D_1)\bar C_2-\bar C_2\ox p_1
                  +2\beta_1\bar C_1+4\beta_0\bar C_2.
 \label{eq:C3-recurrence}
\end{align}
The finite function $c_3(z)$ is an independent physical input. Reversing a product $\bar C_j\ox p_r$ generally changes the result. This is the ordering used for the flavor-matrix validation and for recovering the fixed-scale logarithms from imported coefficients.

\subsection{Contact terms in canonical momentum boundaries}

The generator has the formal plus expansion
\begin{equation}
\begin{aligned}
 \Ggen_\eta
 &=h(\eta)\left[\deltaT+
       \sum_{n\ge0}\frac{(-1)^n\eta^{n+1}}{n!}\Ln_n\right],\\
 \ln h(\eta)&=2\sum_{r\ge1}\frac{\zeta_{2r+1}}{2r+1}\eta^{2r+1}.
\end{aligned}
 \label{eq:generator-plus}
\end{equation}
Integrating over the canonical disk proves eq.~\eqref{eq:boundary-moments}. Differentiation also gives, for example,
\begin{equation}
 \Jdist_5=-6\Ln_5+240\zeta_3\Ln_2+288\zeta_5\Ln_0
                             +160\zeta_3^2\deltaT.
 \label{eq:frak-five}
\end{equation}
The finite canonical momentum CS boundary is consequently
\begin{equation}
 d_{k,i}^{(3)}=d_i^{(3)}-\frac43\zeta_3\beta_0^2\Gamma_0^i.
 \label{eq:CS-disk-shift}
\end{equation}
Equations~\eqref{eq:finite-boundary-shifts} and \eqref{eq:CS-disk-shift} translate to a disk-boundary implementation. The spectral-canonical solution in section~\ref{sec:unexpanded} uses the parent constants $c_n,d_n$ before these shifts. Expanding or integrating that solution in the conventional basis generates the same contact terms automatically.

\section{Weak cumulative profiles and sharp-bin actions}
\label{app:profiles}

This appendix supplies the weak formulation used in section~\ref{sec:fixedorder}. It is useful for comparing fixed-order schemes and for a separately chosen cumulative-profile construction~\cite{EbertTackmann2017}.

For $t=k_T^2$, define
\begin{equation}
\begin{aligned}
 \psi_\varphi(t)&=\frac1{2\pi}\int_0^{2\pi}\dd\phi\,
                \varphi(\sqrt t\,\widehat e_\phi),\\
 C_p(t)&=\Sigma_A(\sqrt t;\lambda_p(\sqrt t)).
\end{aligned}
 \label{eq:profile-data}
\end{equation}
Assume $C_p\in L^1_{\rm loc}([0,\infty))$ with at most polynomial growth. The radial lift of its cumulative derivative is
\begin{equation}
 \pair{\mathcal S_pA}{\varphi}
       =-\int_0^\infty\dd t\,C_p(t)\psi_\varphi'(t),
 \qquad \varphi\in\Sch.
 \label{eq:profile-action}
\end{equation}
For this class the action is a continuous tempered distribution, including finite powers of $\ln t$ at the origin.

\begin{proposition}[Contact-complete cumulative identity]
\label{prop:profile-contact}
At positive continuity points of $C_p$, the common disk and annulus extensions give
\begin{align}
 \int_{|\bk|\le K}\dd^2\bk\,\mathcal S_pA&=C_p(K^2),\notag\\
 \int_{K_a<|\bk|\le K_b}\dd^2\bk\,\mathcal S_pA
       &=C_p(K_b^2)-C_p(K_a^2),\qquad K_a>0.
 \label{eq:profile-bin}
\end{align}
If $C_p(0^+)$ is finite and $C_p-C_p(0^+)\in W^{1,1}_{\rm loc}([0,\infty))$, then
\begin{equation}
 \mathcal S_pA=C_p(0^+)\deltaT+\frac1\pi C_p'(k_T^2).
 \label{eq:profile-split}
\end{equation}
The derivative term is interpreted on $k_T>0$ with the stated local integrability.
\end{proposition}
\begin{proof}
Approximate a disk indicator by smooth radial tests that equal one near $t=0$. Their derivatives tend to $-\delta(t-K^2)$, giving the first identity from eq.~\eqref{eq:profile-action}. Differences give the annulus formula. Under the additional absolute-continuity assumptions, integration by parts gives the endpoint contribution $C_p(0^+)\varphi(0)$ and the regular radial term. For a constant cumulative $C_p=c$ the result is precisely $c\deltaT$. If the origin cumulative diverges logarithmically, the original weak action remains the defining expression.
\end{proof}

For a logarithmic plus distribution,
\begin{equation}
 C_{n,p}(t)=-\frac1{n+1}\ln^{n+1}\frac{\mu_p^2(\sqrt t)}{t},
 \label{eq:log-profile-cumulative}
\end{equation}
whose ordinary density away from the origin is
\begin{equation}
 \frac1{\pi k_T^2}\ln^n\frac{\mu_p^2(k_T)}{k_T^2}
             \left(1-\frac{\dd\ln\mu_p}{\dd\ln k_T}\right).
 \label{eq:log-profile-density}
\end{equation}
This density is supplemented by the contact prescription in eq.~\eqref{eq:profile-action}. A frozen scale has a logarithmically divergent origin cumulative, and is covered by the weak definition. Canonical $\mu_p(K)=K$ gives zero for the logarithmic distributions $\Ln_n$, while the contact pieces of the derivative basis survive. Running coefficients and PDFs are included in the cumulative before differentiation.

For completeness, suppose two cumulative approximations differ by $R_N(t)$ with $|R_N(t)|\le a^{N+1}r(t)$ uniformly on compact intervals containing zero, with locally integrable $r$. For noncompact Schwartz tests require additionally $|R_N(t)|\le a^{N+1}C(1+t)^m$ at large $t$, with constants uniform in the expansion regime. Then
\begin{equation}
 \pair{\mathcal S_p A_N-\mathcal S_p B_N}{\varphi}
       =-\int_0^\infty\dd t\,R_N(t)\psi_\varphi'(t)
       =\Order{a^{N+1}}.
 \label{eq:profile-order}
\end{equation}
This is finite-order inheritance from a dominated weak estimate. The origin condition is essential for first bins. A further all-order logarithmic argument would be needed for a separately proposed resummed profile scheme.

The profile map is linear but generally
\begin{equation}
 \mathcal S_p[A\ot B]\ne(\mathcal S_pA)\ot(\mathcal S_pB).
 \label{eq:profile-nonhomomorphism}
\end{equation}
Its placement in an ordered momentum-space recurrence is therefore part of the prescription. The spectral canonical operator of this paper instead uses the complete RG symbol and its functional calculus; it does not assume eq.~\eqref{eq:profile-nonhomomorphism} can be replaced by an equality.

\section{Input classes, tails, and regulated integral functionals}
\label{app:inputs}

\subsection{Localized smearing and the perturbative tail}

Let $S\in C_c^\infty(\mathbb R^2)$ be radial, normalized to unit integral, and supported inside $|\bp|\le\Lambda_S$. Assume the ordinary perturbative tail is $C^4$ for $k_T>\Lambda_S$. For $k_T>2\Lambda_S$, Taylor expansion with integral remainder gives
\begin{equation}
 [S\ot F^{\pert}](\bk)
 =F^{\pert}(\bk)+\frac14m_2\nabla_k^2F^{\pert}(\bk)+R_4(\bk).
 \label{eq:tail-taylor}
\end{equation}
Radial symmetry removes the odd moments and yields
$\int p_ap_bS(\bp)\dd^2\bp=\delta_{ab}m_2/2$. The remainder is bounded by a constant times
\begin{equation}
 \left(\int\dd^2\bp\,p_T^4|S(\bp)|\right)
 \sup_{|\bp|\le\Lambda_S}\max_{|\alpha|=4}
               |\partial^\alpha F^{\pert}(\bk-\bp)|.
 \label{eq:tail-remainder}
\end{equation}
If the derivatives have the inverse-power/logarithmic bounds of the perturbative tail, the first correction to a $1/k_T^2$-type tail is relatively of order $\Lambda_S^2/k_T^2$, with its accompanying logarithms. The support condition keeps the Taylor argument away from the contact point. This supplies a sufficient model class preserving the leading tail and makes no claim about an unregulated integral of the full perturbative distribution.

A smooth Fourier expansion of this localized input starts with $\widetilde S(\bb)=1-m_2b_T^2/4+\cdots$. Thus its use as the fitted smearing is consistent with a short-distance expansion under the same moment conditions. The evolution and matching operator still fixes the leading perturbative dependence.

\subsection{A sufficient class of broad integral weights}

Let $w_K(\bk)=w(\bk/K)$ with $w\in\Sch$ and $w(0)=1$. Suppose the parent OPE remainder has a uniform small-$b_T$ bound proportional to $(b_T\Lambda)^p$, with any stated logarithmic factors, and controlled polynomial growth outside a fixed small-distance neighborhood. Keep $x$ away from unresummed endpoints and choose scales in the associated perturbative hierarchy. Then its contribution to $\mathcal I_w$ is suppressed by $(\Lambda/K)^p$.

To prove this statement, the dual weight in the Fourier pairing is $K^2\widetilde w(-K\bb)$. Put $\boldsymbol u=K\bb$. In the small-distance region the OPE remainder contributes $(\Lambda/K)^p$ times an integrable moment of the Schwartz function $\widetilde w(\boldsymbol u)$. Outside that region, Schwartz decay suppresses the polynomially bounded remainder by arbitrarily high inverse powers of $K$. The leading coefficient is
\begin{equation}
 Z^w_{i\leftarrow j}(K;\mu,\zeta)
       =\pair{\Ccal_{i\leftarrow j}(\mu,\zeta)}{w_K}.
 \label{eq:Z-weight}
\end{equation}
This proves eq.~\eqref{eq:weighted-OPE} for the specified smooth class. A sharp cutoff or a narrow soft acceptance requires a different endpoint analysis or the full TMD action~\cite{EbertMichelStewartSun2022}.

\section{Endpoint evaluation and finite approximation errors}
\label{app:operator}

For completeness, consider an arbitrary endpoint input $f\in\Dlog$ rather than a canonically preconditioned input. Choose a smooth cutoff $0\le\vartheta_\varepsilon(s)\le1$ that is zero for $s\le\varepsilon$ and one for $s\ge2\varepsilon$, and define $u_\varepsilon=\vartheta_\varepsilon u$. It is bounded, continuous, and constant for sufficiently large $s$. Consequently its compactified continuous function can be approximated by polynomials of $R_\tau$. Since $|u(s)|\le Cw(s)^p$,
\begin{equation}
\begin{aligned}
 \norm{(u-u_\varepsilon)(T)f}_2
 &\le Cw(2\varepsilon)^{-m}\norm{w(T)^{p+m}f}_2\\
 &\longrightarrow0,\qquad m>0 .
\end{aligned}
 \label{eq:endpoint-cutoff}
\end{equation}
Indeed, the difference has spectral support below $2\varepsilon$, where $w(s)\ge w(2\varepsilon)$. Inserting the extra power $w^m$ gives the bound by the spectral theorem. Increasing integer powers can replace noninteger exponents in the domain norm.

For each fixed $\varepsilon$, uniform polynomial convergence gives a momentum-only approximation of $u_\varepsilon(T)$. Equation~\eqref{eq:endpoint-cutoff} then removes the cutoff in a strong limit. This cutoff is a device for evaluation on the declared domain; it is not an additional physical ultraviolet modification of the canonical solution.

The range-space interpretation is also useful. Define
\begin{equation}
 \mathscr H_e=v_i(T;e)L^2,\qquad
 \norm{F}_{\mathscr H_e}=\norm{v_i(T;e)^{-1}F}_2.
 \label{eq:range-space}
\end{equation}
Because $v_i(s;e)>0$ almost everywhere, the inverse on this range is unique. The ratio identity in eq.~\eqref{eq:ratio} makes endpoint evolution an isometry from $\mathscr H_{e_a}$ to $\mathscr H_{e_b}$. This is a domain statement; it does not recommend numerically inverting a smoothing reference operator during a fit.

Finally, for any continuous compactified scalar function $h$, the Bernstein error can be controlled without assuming a finite second derivative. Given $\delta>0$, split its binomial expectation into $|J/N-r|\le\delta$ and the complement. Uniform continuity controls the first part, and the variance bound $r(1-r)/N$ controls the second. If $h\in C^2[0,1]$, Taylor's theorem additionally gives
\begin{equation}
 \norm{B_N[h]-h}_\infty\le\frac{\norm{h''}_\infty}{8N}.
 \label{eq:C2-rate}
\end{equation}
For the physical canonical ultraviolet endpoint, the H\"older bound in theorem~\ref{thm:evaluator} is the generally applicable result under the specified derivative assumptions. A finite-grid spectral comparison controls the approximation on that grid; finite-volume and continuum errors require separate refinement.

\section{Native two-leg reference and ultraviolet completion bound}
\label{app:physical-reference}

\subsection{Algebraic-tail convolution}

For the normalized Student densities in Sec.~\ref{sec:validation}, Feynman
parametrization of the native transverse convolution gives
\begin{align}
 &(f_{p,\lambda_A}\ot f_{q,\lambda_B})(k)\notag\\
 &\quad={\cal N}_{pq}\int_0^1\dd u\,
       \frac{u^{p-1}(1-u)^{q-1}}{D(u,k)^{p+q-1}},\\
 D(u,k)&=u\lambda_A^2+(1-u)\lambda_B^2+u(1-u)k^2,\notag\\
 {\cal N}_{pq}&=
 \frac{(p-1)(q-1)\lambda_A^{2p-2}\lambda_B^{2q-2}}{\pi}
 \frac{\Gamma(p+q-1)}{\Gamma(p)\Gamma(q)}.
 \label{eq:student-convolution}
\end{align}
To obtain this expression, combine the two radial denominators with a Feynman
parameter, shift the two-dimensional integration variable, and use the convergent
integral of $(\ell^2+D)^{-p-q}$. It evaluates the two-leg input without a
transform. Normalization, bin integrals and the logarithmic scale derivative
are checked separately. The continuum reference instead multiplies the two
analytic Bessel-$K$ transforms.

\subsection{A conditional bound above the PDF table}

Let $b_{\rm cut}=b_0/\mu_{\max}$ and $u=\ln(\mu_c/\mu_{\max})$.
Equation~\eqref{eq:pdfuvclass} bounds each PDF by its tabulated endpoint times
$(1+u)^4$. The two-leg luminosity therefore has an eighth-power envelope.
For each strict-NLO component, absolute longitudinal integrals supply a positive
constant $A_{\rm comp}$ that includes flavor/charge sums and bounded coupling
factors. A signed, cancellation-reduced matching value is not substituted for
that absolute envelope.

Monotonic evolution gives a useful UV majorant. Set
$B(a)=\sum_{r=0}^3\beta_r a^{r+2}$, so that $\dd a/\dd\ln\mu=-2B(a)$.
At the symmetric endpoint, differentiating the canonical exponent with respect
to its lower scale gives
\begin{align}
 \frac{\dd\ln v}{\dd\ln\mu_c}
 &=d(a_c)-\gamma(a_c)\notag\\
 &\quad+2\{\Gamma(a_c)+B(a_c)d'(a_c)\}
                  \ln\frac{Q}{\mu_c}.
 \label{eq:UV-monotone}
\end{align}
On the audited interval $0<a_c\le a(\mu_{\max})$,
$(d-\gamma)/a_c\le0$ and $(\Gamma+Bd')/a_c>0$.
These are finite-polynomial sign checks using endpoints and stationary points.
For $\mu_c\ge\mu_{\max}>Q$, $v$ is consequently decreasing.

For a bin $q_a<q_T<q_b$, the inequality
$|qJ_1(qb)|\le q^2b/2$ and the unit bound on the transforms of the normalized
positive inputs give the sufficient component bound
\begin{align}
 |\Delta{\cal B}_{\rm comp}|
 &\le \frac{q_b^2-q_a^2}{2}\,
 b_{\rm cut}^2 v(\mu_{\max})^2 A_{\rm comp}\,I_8,\notag\\
 I_8&=\int_0^\infty\dd u\,e^{-2u}(1+u)^8.
 \label{eq:UV-bound}
\end{align}
The actual calculation sharpens this expression by partitioning $u\in[0,24]$,
using the evolution at each interval's left endpoint, and integrating polynomial
moments analytically. The final interval uses monotonicity for the infinite
remainder. An RG integration margin is included; the calculation is numerical,
not interval arithmetic.

For the $p=3$ Student and Gaussian fixtures, the absolute logarithmic derivative
of the first input has $L^1$ norm at most four. The UV derivative bound therefore
uses four times the ordinary-input envelope. For explicit completed-input
integration, positive Gaussian scale-mixture representations provide an even
tighter bound on the transformed derivative. The class bound and specified-input
integration residual answer different questions and are recorded separately.
The class bound is $1.6683\times10^{-6}$ of the peak, passing the declared
$2\times10^{-6}$ criterion; all primary errors retain that class contribution.

\section{Reproduction and figure provenance}
\label{app:reproduction}

The immutable scientific revision in Ref.~\cite{KSpaceSoftware} contains the
coefficient/PDF source manifests, accepted numerical snapshot and verification
registry. Its portable offline entry point is
\begin{verbatim}
python scripts/run_method_validation.py
\end{verbatim}
Adding \texttt{--physical} and explicit \texttt{--library} and
\texttt{--pdf-data} paths runs the full actual-PDF study, UV completions and
support audit. The reproduction guide supplies resource acquisition, source
hash checks, separate E866/neural commands and plotting instructions.
The runtime and acquired external inputs are kept outside the tracked source.

Figures~\ref{fig:continuum}, \ref{fig:canonicalphysical} and \ref{fig:budget}
are generated from the canonical benchmark records.
Figure~\ref{fig:closure} uses the frozen restricted neural closure.
Table~\ref{tab:physicalaudit} includes the broader UV class errors, rounded upward.
For the neural demonstrator, the linked response-kernel validation is the
refined eight-node mass, sixteen-node longitudinal kernel actually used in
the reported closure and fit. A coarser comparison record is identified
separately. All numerical sources carry hashes; an example's input,
prescription and perturbative content must agree before it is used to support
a claim about this method.

\makeatletter
\immediate\write\@auxout{\string\citation{apsrev42Control}}
\makeatother
\bibliographystyle{apsrev4-2}
\bibliography{references}
\end{document}